\documentclass[%
 reprint,
nofootinbib,
 amsmath,amssymb,
 aps,
twocolumn,
floatfix,
]{revtex4-1}

\usepackage [autostyle, english = american]{csquotes}
\MakeOuterQuote{"}

\usepackage[colorlinks=true]{hyperref}

\hypersetup{colorlinks,citecolor=blue,linkcolor=blue,urlcolor=blue}
\hypersetup{final=true}
\usepackage{aas_macros}
\usepackage{todonotes}
\usepackage{graphicx}
\usepackage{caption}
\usepackage{subcaption}

\usepackage{bm}

\begin{document}


\title{Einstein's Universe: Cosmological structure formation in numerical relativity}

\author{Hayley J. Macpherson}
  \email{hayley.macpherson@monash.edu}
 \affiliation{%
 Monash Centre for Astrophysics and School of Physics and Astronomy, Monash University, VIC 3800, Australia}

\author{Daniel J. Price}
\affiliation{%
Monash Centre for Astrophysics and School of Physics and Astronomy, Monash University, VIC 3800, Australia}%

\author{Paul D. Lasky}
\affiliation{%
Monash Centre for Astrophysics and School of Physics and Astronomy, Monash University, VIC 3800, Australia}%
\affiliation{%
OzGrav: The ARC Centre of Excellence for Gravitational-wave Discovery, Clayton, VIC 3800, Australia}%

\date{\today}

\begin{abstract}
We perform large-scale cosmological simulations that solve Einstein's equations directly via numerical relativity. Starting with initial conditions sampled from the cosmic microwave background, we track the emergence of a cosmic web without the need for a background cosmology. We measure the backreaction of large-scale structure on the evolution of averaged quantities in a matter-dominated universe. Although our results are preliminary, we find the global backreaction energy density is of order $10^{-8}$ compared to the energy density of matter in our simulations, and is thus unlikely to explain accelerating expansion under our assumptions. Sampling scales above the homogeneity scale of the Universe ($100-180\,h^{-1}$Mpc), in our chosen gauge, we find $2-3\%$ variations in local spatial curvature.


\end{abstract}

\pacs{Valid PACS appear here}
\maketitle




%
%

\section{Introduction}
Modern cosmology derives from the Friedman-Lema\^itre-Robertson-Walker (FLRW) metric --- an exact solution to Einstein's equations that assumes homogeneity and isotropy. The formation of cosmological structure means that the Universe is neither homogeneous nor isotropic on small scales. The Lambda Cold Dark Matter ($\Lambda$CDM) model assumes the FLRW metric, and has been the leading cosmological model since the discovery of the accelerating expansion of the Universe \citep{riess1998,perlmutter1999}. Since then it has had many successful predictions, including the location of the baryon acoustic peak \citep[e.g.][]{kovac2002,eisenstein2005,cole2005,blake2011a,ata2018}, the polarisation of the cosmic microwave background (CMB) \citep{planck2016params,hinshaw2013}, galaxy clustering, and gravitational lensing \citep[e.g.][]{bonvin2017,hildebrandt2017,DESCollab2017a}. Despite these successes, tensions with observations have arisen. Most notable is the recent $3.8\sigma$ tension between measurements of the Hubble parameter, $H_{0}$, locally \citep{riess2018b} and the value inferred from the CMB under $\Lambda$CDM \citep{planck2016params}. 

The assumptions underlying the standard cosmological model are based on observations that our Universe is, \textit{on average}, homogeneous and isotropic. However, the averaged evolution of an inhomogeneous universe does not coincide with the evolution of a homogeneous universe \citep{buchertehlers1997,buchert2000a}. Additional "backreaction" terms exist, but their significance has been debated \citep[e.g.][]{rasanen2006a,rasanen2006b,li2007,li2008,larena2009,clarkson2011a,wiltshire2011,wiegand2012,green2012,buchert2012,green2014,buchert2015,green2015,bolejko2017d,roukema2017,kaiser2017,buchert2018a}.

State-of-the-art cosmological simulations currently employ the FLRW solution coupled with a Newtonian approximation for gravity \citep{springel2005,kim2011,genel2014}. These simulations have proven extremely valuable to furthering our understanding of the Universe. However, general relativistic effects on our observations cannot be fully studied when the formation of large-scale structure has no effect on the surrounding spacetime. Whether or not these effects are significant can only be tested with numerical relativity, which allows us to fully remove the assumptions of homogeneity and isotropy. Initial works have shown emerging relativistic effects such as differential expansion \citep{bentivegna2016a}, variations in proper length and luminosity distance relative to FLRW \citep{giblin2016a,giblin2016b}, and the emergence of tensor modes and gravitational slip \citep{macpherson2017a}. A comparison between Newtonian and fully general relativistic simulations found sub-percent differences within the weak-field regime \citep{east2018}, in agreement with post-Friedmannian N-body calculations \citep{adamek2013,adamek2014b}. 

In this work, we present cosmological simulations with numerical relativity, using realistic initial conditions, evolved over the entire history of the Universe. Here we use a fluid approximation for dark matter, however, this is one more step along the road to fully relativistic cosmological N-body calculations. We focus on the global backreaction of cosmological structures on averaged quantities, including the matter, curvature, and backreaction energy densities, and how these averages vary as a function of physical size of the averaging domain. We test the global and local effects on the expansion rate, including the potential for backreaction to contribute to the accelerating expansion of the Universe. In a companion paper we examine whether local variations in the Hubble expansion rate can explain the discrepancy between local and global measurements of the Hubble constant \citep{macpherson2018b}.



In Section~\ref{sec:compsetup} we describe our computational setup, in Section~\ref{sec:ics} we describe the derivation and implementation of initial conditions drawn from the CMB, in Section~\ref{sec:gauge} and \ref{sec:averaging} we describe our choice of gauge and averaging scheme respectively, and in Section~\ref{sec:results} we present our simulations and averaged quantities. We discuss our results in Section~\ref{sec:discussion} and conclude in Section~\ref{sec:conclude}. Unless otherwise stated, we adopt geometric units with $G=c=1$, where $G$ is the gravitational constant and $c$ is the speed of light. Greek indices take values 0 to 3, and Latin indices from 1 to 3, with repeated indices implying summation.


\section{Computational Setup} \label{sec:compsetup}
\subsection{Cactus and FLRWSolver}
To evolve a fully general relativistic cosmology we use the open-source Einstein Toolkit \citep{loffler2012}, a collection of codes based on the Cactus framework \citep{cactus}. Within this toolkit we use the \texttt{ML\_BSSN} thorn \citep{brownD2009} for evolution of the spacetime variables using the BSSN formalism \citep{shibata1995,baumgarte1999}, and the \texttt{GRHydro} thorn for evolution of the hydrodynamics \citep{baiotti2005,giacomazzo2007,mosta2014}. In addition, we use our initial-condition thorn, \texttt{FLRWSolver} \citep{macpherson2017a}, to initialise linearly-perturbed FLRW spacetimes with perturbations of either single-mode or CMB-like distributions.

We assume a dust universe, implying pressure $P=0$, however \texttt{GRHydro} currently has no way to implement zero pressure for hydrodynamical evolution. Instead we set $P\ll\rho$, with a polytropic equation of state,
\begin{equation}
	P = K_\mathrm{poly} \rho\,^{2},
\end{equation}
where $K_\mathrm{poly}$ is the polytropic constant, which we set $K_\mathrm{poly}=0.1$ in code units. We have found this to be sufficient to match the evolution of a homogeneous, isotropic, matter-dominated universe. Deviations from the exact solution for the scale factor evolution, at $80^{3}$ resolution, are within $10^{-6}$ \citep[see][]{macpherson2017a}. 

We perform a series of simulations with varying resolutions, $64^{3}, 128^{3}$, and $256^{3}$, and comoving physical domain sizes, $L=100$ Mpc, 500 Mpc, and 1 Gpc, to study different physical scales. We simulate all three domain sizes at $64^{3}$ and $128^{3}$ resolution, and only the $L=1$ Gpc domain size at $256^3$ resolution due to computational constraints. During the evolution we do not assume a cosmological background, and for convenience, since we have not yet implemented a cosmological constant in the Einstein Toolkit, we assume $\Lambda=0$.

Post-processing analysis is performed using the \texttt{mescaline} code, which we introduce and describe in Section~\ref{subsec:mescaline}.

\subsection{Length unit}
We choose the comoving length unit of our simulation domain to be 1 Mpc, implying a domain of $L=100$ in code units is equivalently $L=100$ Mpc. 
In geometric units $c=1$, and so we can relate our length unit, $l=1$ Mpc, and our time unit, $t_c$, via the speed of light (in physical units)
\begin{align} \label{eq:tunit}
	\frac{l}{t_c} &= c = 3\times10^8 \,\mathrm{m\,s}^{-1}.
\end{align}
To find our background FLRW density we use $H(z=0)=H_0$, with units of s$^{-1}$. This implies
\begin{equation} \label{eq:H0units}
H_{0,\mathrm{code}} \times \frac{1}{t_c} = H_{0,\mathrm{phys}},
\end{equation}
where $H_{0,\mathrm{code}}$ and $H_{0,\mathrm{phys}}=100\,h\,\mathrm{km\,s^{-1}\,Mpc^{-1}}$ are the Hubble parameter expressed in code units and physical units, respectively. We use \eqref{eq:tunit} together with \eqref{eq:H0units} and the Friedmann equation for a flat, matter-dominated model
\begin{equation}
	H = \frac{\dot{a}}{a} = \sqrt{\frac{8\pi G\bar{\rho}}{3}},
\end{equation}
where an overdot represents a derivative with respect to proper time, $\bar{\rho}$ is the homogeneous density, and $a$ is the FLRW scale factor. We find the background FLRW density, evaluated at $z=0$, in code units, to be
\begin{equation} \label{eq:rho0_a01}
	\bar{\rho}_{0,\mathrm{code}} = 1.328\times10^{-8}\,h^{2}.
\end{equation}

For computational reasons we adopt the initial FLRW scale factor $a_\mathrm{init}=a(z=1100)=1$, whilst the usual convention in cosmology is to set $a_0=a(z=0)=1$. The density \eqref{eq:rho0_a01} was calculated using the Hubble parameter $H_{0,\mathrm{phys}}$ evaluated with $a_0=1$. The comoving (constant) FLRW density is $\rho^*=\bar{\rho}\,a^3=\bar{\rho}_0\,a_0^3$, and so \eqref{eq:rho0_a01} is the comoving density $\rho^*$. We choose $h=0.704$, and our choice $a_\mathrm{init}=1$ implies our initial background density is the comoving FLRW density.

\subsection{Redshifts}
Simulations are initiated at $z=1100$ and evolve to $z=0$. We quote redshifts computed from the value of the FLRW scale factor at a particular conformal time,
\begin{equation}
	a(\eta) = \frac{z_\mathrm{cmb} + 1}{z(\eta) + 1},
\end{equation}
where $z_\mathrm{cmb}=1100$. Since we set $a_\mathrm{init}=1$, we have $a_0=1101$. The evolution of the FLRW scale factor in conformal time is
\begin{equation}
	a(\eta) = a_\mathrm{init} \, \xi^2, \label{eq:confa}
\end{equation}
where $\xi$ is the scaled conformal time defined in Section~\ref{subsec:linperturb}. Importantly, the redshifts presented throughout this paper are indicative only of the amount of coordinate time that has passed, and are not necessarily indicative of redshifts measured by observers in an inhomogeneous universe.


\section{Initial conditions} \label{sec:ics}

\subsection{Linear Perturbations} \label{subsec:linperturb}
We solve the linearly-perturbed Einstein equations to generate our initial conditions. Assuming only scalar perturbations, the linearly-perturbed FLRW metric in the longitudinal gauge is
\begin{equation} \label{eq:longitudinal_metric}
	{\rm d}s^{2} = -a^{2}(\eta)\left(1+2\psi\right){\rm d}\eta^{2} + a^{2}(\eta)\left(1-2\phi\right){\rm d}x^{i}{\rm d}x^{j}\delta_{ij}.
\end{equation}
In this gauge the metric perturbations $\phi$ and $\psi$ are the Bardeen potentials \citep{bardeen1980}. These are related to perturbations in the matter distribution via the linearly perturbed Einstein equations
\begin{equation}\label{eq:perturbed_einstein}
	\bar{G}_{\mu\nu} + \delta G_{\mu\nu} = 8\pi\left(\bar{T}_{\mu\nu} + \delta T_{\mu\nu}\right),
\end{equation}
where an over-bar represents a background quantity, and $\delta X$ represents a small perturbation in the quantity $X$, with $\delta X\ll X$. A matter-dominated (dust) universe has stress-energy tensor
\begin{equation}
	T_{\mu\nu} = \rho \,u_{\mu}u_{\nu},
\end{equation}
where $\rho$ is the rest-mass density, $u^{\mu}=dx^{\mu}/d\tau$ is the four-velocity of the fluid, and $\tau$ is the proper time. Assuming small perturbations to the matter we have
\begin{align}
	\rho &= \bar{\rho} + \delta\rho = \bar{\rho}(1+\delta), \\
	v^{i} &= \delta v^{i},
\end{align}
where the fractional density perturbation is $\delta\equiv\delta\rho/\bar{\rho}$, and $v^{i}=dx^{i}/d\eta$ is the three-velocity.

Solutions to \eqref{eq:perturbed_einstein} are found by taking the time-time, time-space, trace and trace-free components, given by
 \begin{subequations} \label{eqs:perturbed_einstein}
	\begin{align}
		\nabla^{2}\phi - 3 \mathcal{H}\left(\phi' + \mathcal{H} \psi\right) &= 4\pi  \bar{\rho}\,\delta a^{2}, \label{eq:einstein_1} \\ 
		\mathcal{H} \partial_{i}\psi + \partial_{i}\phi' &= -4\pi \bar{\rho} \,a^{2} \delta_{ij}v^{j}, \label{eq:einstein_2} \\ 
		\phi'' + \mathcal{H}\left(\psi' + 2\phi' \right) &= \frac{1}{2}\nabla^{2}(\phi - \psi), \label{eq:einstein_3} \\ 
		\partial_{\langle i}\partial_{j\rangle} \left(\phi - \psi\right) &= 0, \label{eq:einstein_4}
	\end{align}	
\end{subequations}
respectively, where we have assumed all perturbations are small such that second-order (and higher) terms can be neglected. Here, $\partial_{i} \equiv \partial / \partial x^{i}$, $\nabla^{2}=\partial^{i}\partial_{i}$, $\partial_{\langle i}\partial_{j\rangle}\equiv \partial_{i}\partial_{j} - 1/3\,\delta_{ij}\nabla^{2}$, a $'$ represents a derivative with respect to conformal time, and $\mathcal{H}\equiv a'/a$ is the conformal Hubble parameter. Solving these equations, we find
\begin{subequations} \label{eqs:linear_solns}
    \begin{align}
    	\psi &= \phi = f(x^{i}) - \frac{g(x^{i})}{5\,\xi^{5}}, \\
     	\delta &= C_{1}\, \xi^{2}\, \nabla^{2}f(x^{i}) - 2 \,f(x^{i}) - C_{2} \,\xi^{-3}\,\nabla^{2}g(x^{i}) - \frac{3}{5} \xi^{-5} g(x^{i}), \\
     	v^{i} &= C_{3}\,\xi\, \partial^{i}f(x^{i}) + \frac{3}{10}C_{3}\,\xi^{-4}\, \partial^{i}g(x^{i}),
    \end{align}
\end{subequations}
where $f, g$ are arbitrary functions of spatial position, we introduce the scaled conformal time coordinate
\begin{equation}
	\xi \equiv 1 + \sqrt{\frac{2\pi\rho^{*}}{3\,a_\mathrm{init}}}\eta,
\end{equation}
and we have defined
\begin{equation} 
	C_{1}\equiv \frac{a_{\mathrm{init}}}{4\pi\rho^{*}},\quad C_{2}\equiv \frac{a_{\mathrm{init}}}{20\pi\rho^{*}}, \quad C_{3}\equiv-\sqrt{\frac{a_{\mathrm{init}}}{6\pi\rho^{*}}}.
\end{equation}
Equations \eqref{eqs:linear_solns} contain both a growing and decaying mode for the density and velocity perturbations. We choose $g=0$ to extract only the growing mode of the density perturbation, and our solutions become
\begin{subequations} \label{eqs:linear_solnsg0}
    \begin{align}
    	\psi &= \phi = f(x^{i}), \\
     	\delta &= C_{1}\, \xi^{2}\, \nabla^{2}f(x^{i}) - 2 \,f(x^{i}), \\
     	v^{i} &= C_{3}\,\xi\, \partial^{i}f(x^{i}),
    \end{align}
\end{subequations}
implying $\phi'=0$ in the linear regime.


\begin{figure}
	\includegraphics[width=0.5\textwidth]{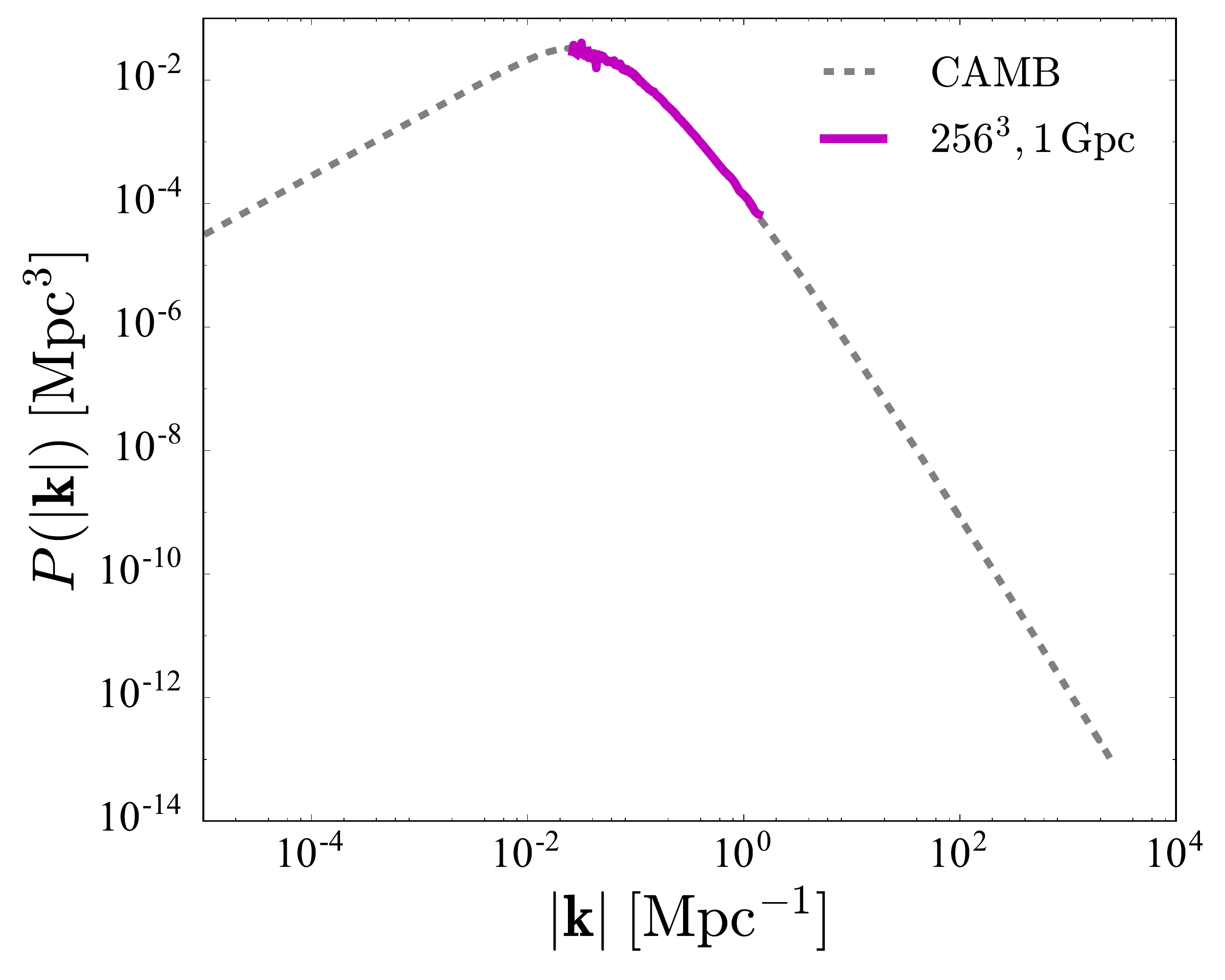}
    \caption{Matter power spectrum of our initial conditions. Grey dashed curve shows the power spectrum produced with the Code for Anisotropies in the Microwave Background (CAMB). We show the power as a function of wavenumber $|\textbf{k}|=\sqrt{k_x^2+k_y^2+k_z^2}$. The magenta curve shows the section of the power spectrum we sample when using a domain size of $L=1$ Gpc with resolution $256^{3}$.}
    \label{fig:powerspectrum}
\end{figure}
\begin{figure*}
	\begin{center}
	\includegraphics[width=\textwidth]{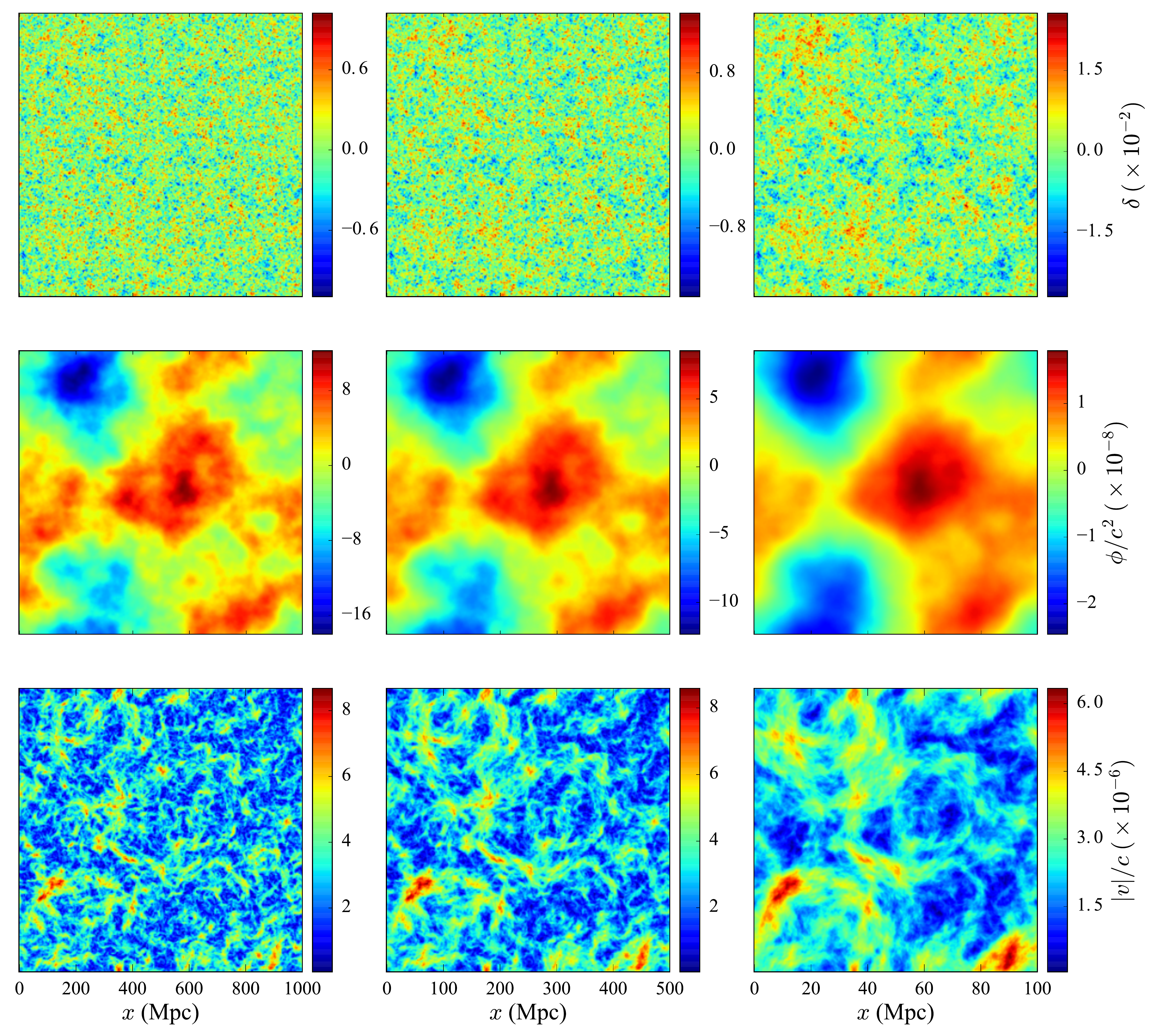}
    \caption{Initial conditions drawn from the cosmic microwave background power spectrum. Here we show initial conditions for the density (top row), metric (middle row), and velocity (bottom row) perturbations for three different physical domain sizes. Left to right shows domain sizes $L=1$ Gpc, 500 Mpc, and 100 Mpc. We show a two-dimensional slice through the midplane of each domain. All initial conditions shown here are at $256^{3}$ resolution, and all quantities are shown in code units -- normalised by the speed of light for the metric and velocity perturbations. The magnitude of the velocity is $|v|=\sqrt{v_x^{2} + v_y^{2} + v_z^{2}}$.}
    \label{fig:256ICs}
	\end{center}
\end{figure*}
\subsection{Cosmic Microwave Background fluctuations} \label{subsec:cmbfluc}
We use \eqref{eqs:linear_solns} along with the Code for Anisotropies in the Microwave Background (CAMB) \citep{lewis2002} to generate the matter power spectrum at $z=1100$, with parameters consistent with \citet{planck2016params} as input. Figure~\ref{fig:powerspectrum} shows the matter power spectrum from CAMB (grey curve), as a function of wavenumber $|\textbf{k}|=\sqrt{k_x^{2} + k_y^{2} + k_z^{2}}$. We use the Python module \texttt{c2raytools} \footnote{https://github.com/hjens/c2raytools} to generate a 3-dimensional Gaussian random field drawn from the CAMB power spectrum. This provides the initial density perturbation. The magenta curve in Figure~\ref{fig:powerspectrum} shows the region of the matter power spectrum sampled in our highest resolution ($256^{3}$), largest domain size ($L=1$ Gpc) simulation. The smallest $\textbf{k}$ component sampled represents the largest wavelength of perturbations --- approximately the length of the box, $L$ --- and the largest $\textbf{k}$ component sampled represents the smallest wavelength of perturbations --- two grid points. To relate the initial density perturbation to the corresponding velocity and metric perturbations, we transform \eqref{eqs:linear_solns} into Fourier space. Initially, $\xi=1$ which gives a density perturbation of the form
\begin{equation}
	\delta(\textbf{k}) = -\left( C_1 |\textbf{k}|^{2} + 2 \right) \,\phi(\textbf{k}),
\end{equation}
where $\textbf{k}=(k_x, k_y, k_z)$, so we can define an arbitrary function $\delta(\textbf{k})$, and construct the metric perturbation and velocity, respectively, using
\begin{subequations}\label{eqs:Fspace}
    \begin{align} 
    	\phi(\textbf{k}) &= -\frac{\delta(\textbf{k})}{C_1 |\textbf{k}|^{2} + 2}, \\
    	\textbf{v}(\textbf{k}) &= C_3 \,i\,\textbf{k}\,\phi(\textbf{k}),
    \end{align}
\end{subequations}
where $i^{2}=-1$. With the Fourier transform of the Gaussian random field as $\delta(\textbf{k})$, we calculate the velocity and metric perturbations in Fourier space using \eqref{eqs:Fspace}, and then use an inverse Fourier transform to convert the perturbations to real space. The density perturbation $\delta$ is already dimensionless, and we normalise by the speed of light, $c$, to convert $v^{i}$ and $\phi$ to code units. Figure~\ref{fig:256ICs} shows initial conditions at $256^{3}$ resolution for box sizes $L=1$ Gpc, 500 Mpc, and 100 Mpc in the left to right columns, respectively. The top row shows the density perturbation $\delta$, the middle row shows the normalised metric perturbation $\phi/c^2$, and the bottom row shows the magnitude of the velocity perturbation normalised to the speed of light $|v|/c$. These initial conditions are sufficient to describe a linearly-perturbed FLRW spacetime in \texttt{FLRWSolver}.

We assume a flat FLRW cosmology for the initial instance only. Simulations begin with small perturbations at the CMB, and so the assumption of a linearly-perturbed FLRW spacetime is sufficiently accurate.

\section{Gauge} \label{sec:gauge}
The (3+1) decomposition of Einstein's equations \citep{arnowitt1959} results in the metric
\begin{equation} \label{eq:3p1metric}
	{\rm d}s^{2} = -\alpha^{2}{\rm d}t^{2} + \gamma_{ij}\left( {\rm d}x^{i} + \beta^{i}{\rm d}t \right)\left({\rm d}x^{j} + \beta^{j}{\rm d}t\right),
\end{equation}
where $\gamma_{ij}$ is the spatial metric, $\alpha$ is the lapse function, $\beta_{i}$ is the shift vector, $x^{i}$ are the spatial coordinates, and $t$ is the coordinate time. The lapse function determines the relationship between proper time and coordinate time from one spatial slice to the next, while the shift vector determines how spatial points are relabelled between slices. In cosmological simulations with numerical relativity the comoving synchronous gauge (geodesic slicing) is a popular choice \citep[e.g.][]{bentivegna2016a,giblin2016a,giblin2016b,giblin2017a,giblin2017b}, which involves fixing $\alpha=1$, $\beta_{i}=0$, and $u^{\mu}=(1,0,0,0)$, or $u^{\mu}=(1/a,0,0,0)$ for conformal time, throughout the simulation. This gauge choice can become problematic at low redshifts when geodesics begin to cross, and can form singularities. 
We choose $\beta_{i}=0$ and evolve the lapse according to the general spacetime foliation
\begin{equation} \label{eq:lapseevol}
	\partial_{t}\alpha = - f(\alpha) \,\alpha^{2} K,
\end{equation}
where $f(\alpha)$ is a positive and arbitrary function, and $K=\gamma^{ij}K_{ij}$ is the trace of the extrinsic curvature.  We choose $f=1/3$, and use the relation from the (3+1) ADM equations \citep{shibata1995}
\begin{equation}
	\partial_{t} \,\mathrm{ln}(\gamma^{1/2}) = -\alpha K,
\end{equation}
where $\gamma$ is the determinant of the spatial metric. Integrating \eqref{eq:lapseevol} gives
\begin{equation}
	\alpha = C(x^{i})\,\gamma^{1/6},
\end{equation}
where $C(x^{i})$ is an arbitrary function of spatial position.

For our initial conditions we have $\gamma_{ij}=a^{2}(1-2\phi)\delta_{ij}$, implying $\gamma^{1/6}=a\,\sqrt{1-2\phi}$. We therefore choose
\begin{equation}
	C(x^{i}) = \frac{\sqrt{1+2\psi}}{\sqrt{1-2\phi}}
\end{equation}
on the initial hypersurface, so that $\alpha=a\,\sqrt{1+2\psi}$, as in the metric \eqref{eq:longitudinal_metric}. 

\section{Averaging scheme} \label{sec:averaging}
We adopt the averaging scheme of \citet{buchert2000a} generalised for an arbitrary coordinate system \citep{larena2009b,brown2009a,brown2009b,clarkson2009,gasperini2010,umeh2011}  \footnote{During the review of this paper, \citet{buchert2018b} raised some concerns regarding the averaging formalism of \citet{larena2009b} and \citet{brown2009b}. We aim to investigate the proposed alterations in a later work.}. The average of a scalar quantity $\psi(x^{i},t)$ is defined as
\begin{equation}\label{eq:avgdef}
	\langle\psi\rangle = \frac{1}{V_{\mathcal{D}}} \int_{\mathcal{D}}\psi \sqrt{\gamma}\;{\rm d}^{3}X,
\end{equation}
where the average is taken over some domain $\mathcal{D}$ lying within the chosen hypersurface, and $V_{\mathcal{D}} = \int_{\mathcal{D}} \sqrt{\gamma} d^{3}X$
 is the volume of that domain. The normal vector to our averaging hypersurface is $n_{\mu}=(-\alpha,0,0,0)$, corresponding to the four-velocity of observers within our simulations. These  observers are not comoving with the fluid, implying $n_\mu \neq u_{\mu}$, and the tilt between these two vectors results in additional backreaction terms due to nonzero peculiar velocity $v^{i}$. As in \citep{larena2009b,clarkson2009,brown2013}, we define the Hubble expansion of a domain $\mathcal{D}$ to be associated with the expansion of the fluid, $\theta$,
 \begin{equation} \label{eq:hubble_def}
 	\mathcal{H_{D}} \equiv \frac{1}{3}\langle \theta \rangle,
 \end{equation}
 where       
 \begin{equation} \label{eq:theta}
 	\theta\equiv h^{\alpha\beta}\nabla_{\alpha}u_{\beta},
\end{equation}
is the projection of the fluid expansion onto the three-surface of averaging, with the projection tensor $h_{\alpha\beta} \equiv g_{\alpha\beta} + n_{\alpha}n_{\beta}$. In our case, this represents the expansion of the fluid as observed in the gravitational rest frame \citep{umeh2011}. 

Averaging Einstein's equations in this frame, with $P=\Lambda=0$, gives the averaged Hamiltonian constraint
\begin{equation} \label{eq:buchert1}
	6\mathcal{H_D}^{2} = 16\pi\langle\Gamma^{4}\rho\rangle - \mathcal{R_D} - Q_\mathcal{D} + \mathcal{L_D},
\end{equation}
where $\Gamma$ is the Lorentz factor, $\mathcal{R_D}$ is the averaged Ricci curvature scalar, $Q_\mathcal{D}$ is the dynamical backreaction term, and $\mathcal{L_D}$ is the additional backreaction term due to nonzero peculiar velocities in our gauge. For definitions of these terms, see Appendix~\ref{sec:appx_avg}.

\begin{figure*}
	\includegraphics[width=\textwidth]{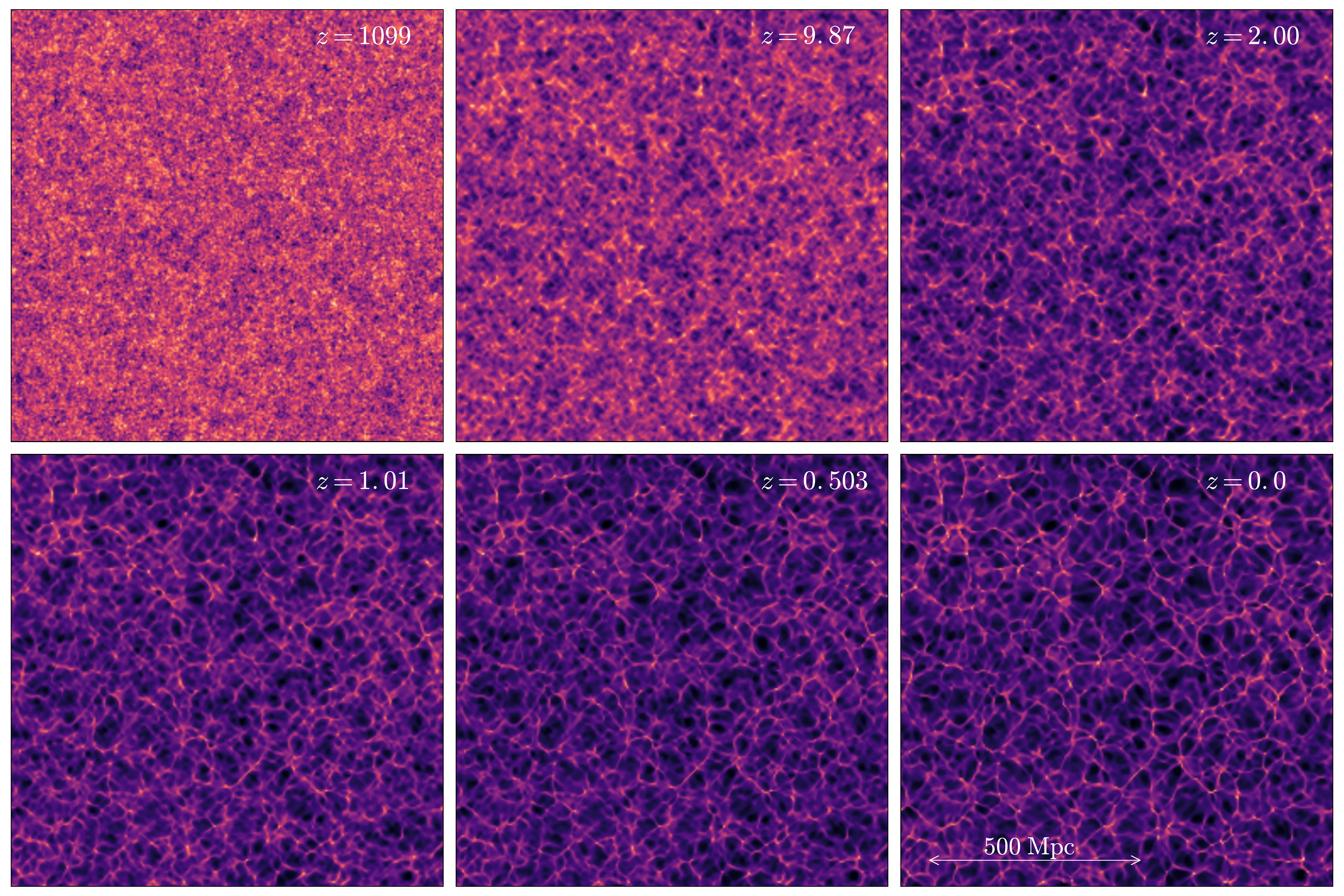}
    \caption{Evolution of a fully general-relativistic cosmic web. Here we show a $256^{3}$ simulation, in an $L=1$ Gpc domain. This simulation has evolved from the cosmic microwave background ($z=1100$; top left) until today ($z=0$; bottom right). Each panel shows a two-dimensional slice of the density perturbation in the midplane of the domain. We can see the familiar web structure of modern cosmological N-body simulations using Newtonian gravity, however this cosmic web contains all of the corresponding general relativistic information. The standard deviations of the fractional density perturbation $\delta$ for each panel (progressing in time) are $\sigma_{\delta}=0.0026, 0.15, 0.6, 1.11, 1.89$, and $3.92$, respectively.}
    \label{fig:256evol}
\end{figure*}

We define the effective scale factor, $a_\mathcal{D}$, describing the expansion of the fluid, via the Hubble parameter
 \begin{equation} \label{eq:aDdef}
 	\mathcal{H_{D}} = \frac{a'_{\mathcal{D}}}{a_{\mathcal{D}}}.
\end{equation}
This is related to the effective scale factor describing the expansion of the coordinate grid (volume)
\begin{equation} \label{eq:aDVdef}
	a_{\mathcal{D}}^{V}\equiv \frac{V'_\mathcal{D}}{V_\mathcal{D}} = \left(\frac{V_\mathcal{D}(\eta)}{V_\mathcal{D}(\eta_\mathrm{init})}\right)^{1/3},
\end{equation}
via 
\begin{equation}\label{eq:aDaDV}
	a_{\mathcal{D}} = a_{\mathcal{D}}^{V} \;\mathrm{exp}\left(-\frac{1}{3}\int \langle \,\alpha\;\Gamma^{-1}\;(\theta - \kappa) - \alpha\,\theta\,\rangle \;{\rm d}\eta\right).
\end{equation}
See Appendix~\ref{sec:appx_expn} for details.

\subsection{Cosmological parameters}
The dimensionless cosmological parameters describe the content of the Universe. From \eqref{eq:buchert1} we define
\begin{subequations} \label{eqs:cosmic_qrtet}
	\begin{align}
		\Omega_{m} &= \frac{8\pi \langle\Gamma^{2}\rho\rangle}{3 \mathcal{H_{D}}^{2}}, \quad \Omega_{R} = -\frac{\mathcal{R_{D}}}{6 \mathcal{H_{D}}^{2}}, \\
		 \Omega_{Q} &= - \frac{Q_{\mathcal{D}}}{6 \mathcal{H_{D}}^{2}}, \quad \Omega_{L} = \frac{\mathcal{L_{D}}}{6\mathcal{H_{D}}^{2}},
	\end{align}
\end{subequations}
giving the Hamiltonian constraint in the form
\begin{equation} \label{eq:content}
	\Omega_{m} + \Omega_{R} + \Omega_{Q} + \Omega_{L} = 1.
\end{equation}
We require this to be satisfied at all times. Here, $\Omega_m$ is the matter energy density, $\Omega_R$ is the curvature energy density, $\Omega_Q+\Omega_L$ is the energy density associated with the backreaction terms; a purely general relativistic effect. For a standard $\Lambda$CDM cosmology, these cosmological parameters are $\Omega_m=0.308\pm0.012$, $|\Omega_R|=|\Omega_k|<0.005$, $\Omega_{Q}=0$, and $\Omega_{L}=0$ \citep{planck2016params}.

\begin{figure*}
	\includegraphics[width=\textwidth]{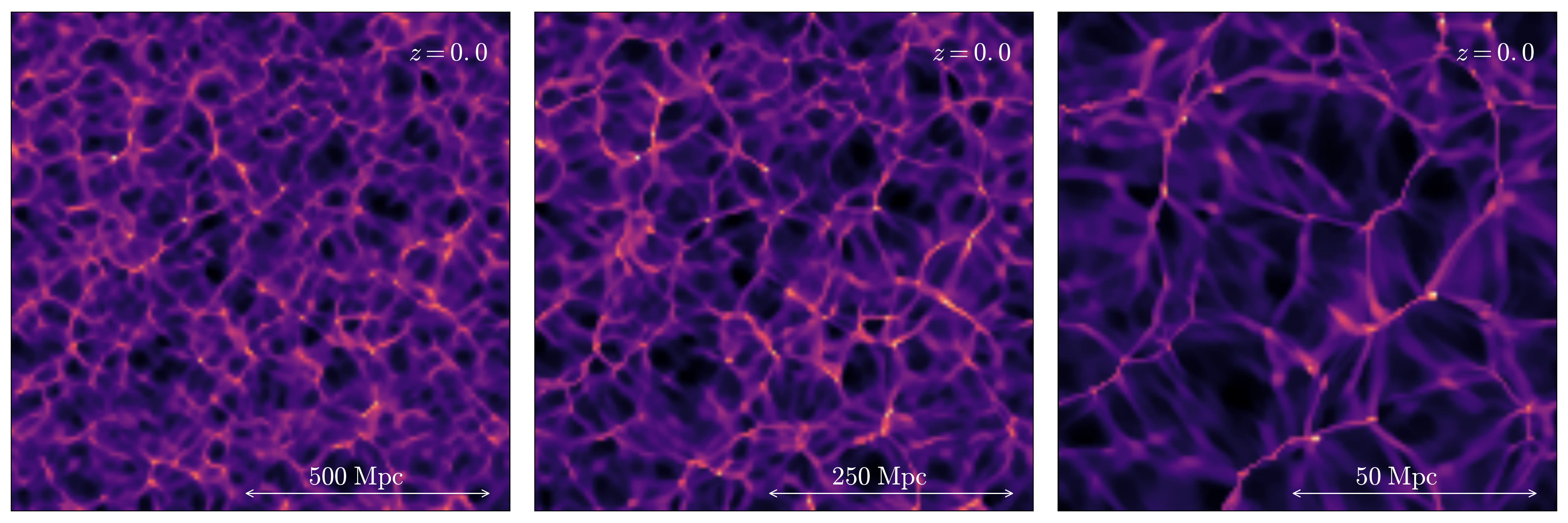}
    \caption{Scale dependence of the cosmic web. Three separate simulations computed at a resolution of $128^{3}$ (left to right) with domain sizes $L=1$ Gpc, 500 Mpc, and 100 Mpc, respectively. All snapshots show a two-dimensional density slice in the midplane of the simulation domain at redshift $z=0$.}
    \label{fig:128allsim_z0}
\end{figure*}
\begin{figure*}
	\includegraphics[width=\textwidth]{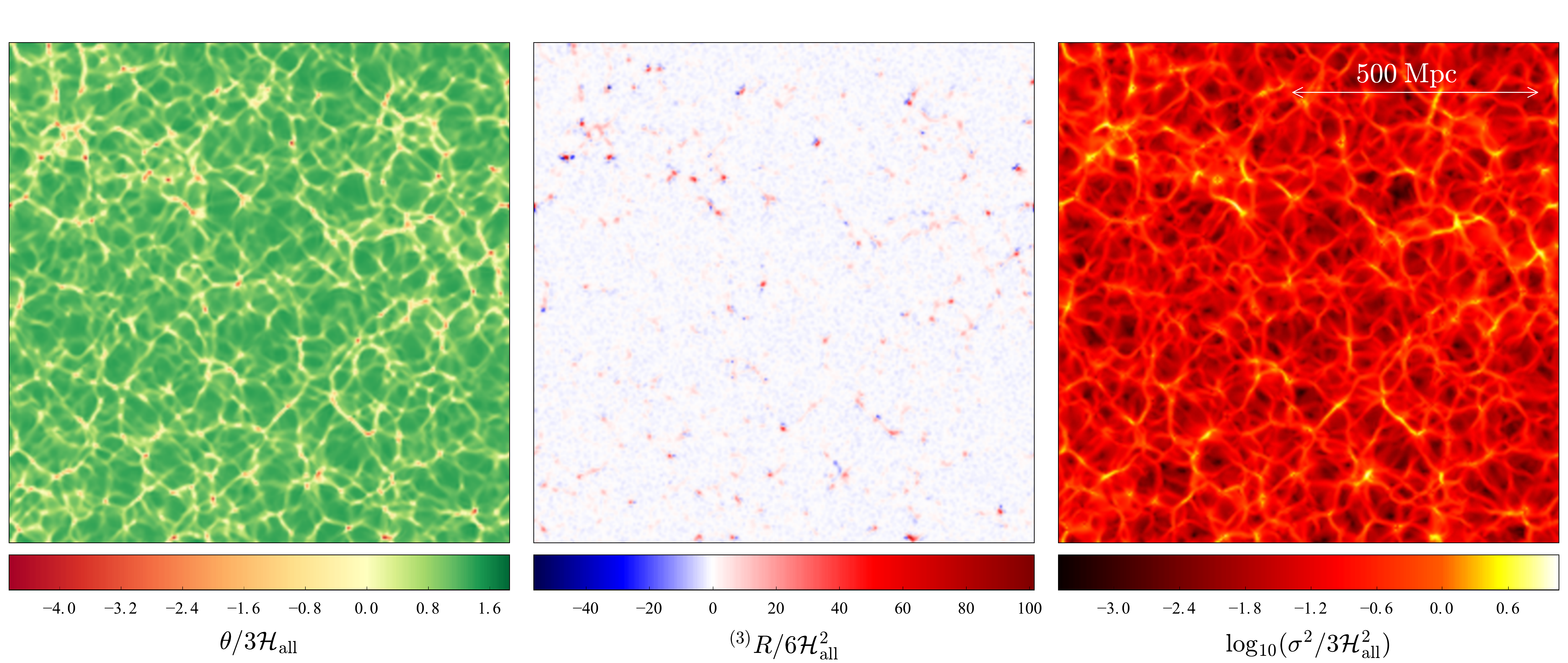} 
    \caption{General relativistic attributes of an inhomogeneous, anisotropic universe. Panels (left to right) show the matter expansion rate $\theta$, the spatial Ricci curvature $R$, and the shear $\sigma^{2}$, respectively, each relative to the global Hubble expansion $\mathcal{H}_{\rm all}$. Each panel shows a two-dimensional slice at $z=0$ through the midplane of the $L=1$ Gpc domain at $256^{3}$ resolution.}
    \label{fig:GRstuff}
\end{figure*}

\subsection{Post-simulation analysis} \label{subsec:mescaline}
The Universe is measured to be homogeneous and isotropic on scales larger than $\sim 80-100 h^{-1} \mathrm{Mpc}$ \cite{scrimgeour2012}. Above these scales it is unclear whether the evolution of the average of our inhomogeneous Universe coincides with the FLRW (or $\Lambda$CDM) equivalent. In attempt to address this, we calculate averages over our entire simulation domain, but also over subdomains within the simulation to sample a variety of physical scales. We measure averages over spheres of varying radius $r_{\mathcal{D}}$ embedded in the total volume, from which we calculate the dimensionless cosmological parameters \eqref{eqs:cosmic_qrtet}, the Hubble parameter \eqref{eq:hubble_def}, and consequently the effective matter expansion $a_\mathcal{D}$.

The spatial Ricci tensor $R_{ij}$ is the contraction of the Riemann tensor. We calculate this directly from the metric using
\begin{equation} \label{eq:ricci4Ddef}
	R_{ij} = \partial_k \Gamma^{k}_{ij} - \partial_j\Gamma^{k}_{ik} + \Gamma^{k}_{lk}\Gamma^{l}_{ij} - \Gamma^{k}_{jl}\Gamma^{l}_{ik},
\end{equation}
where the spatial connection coefficients are
\begin{equation} \label{eq:christoffeldef}
	\Gamma^{k}_{ij} \equiv \frac{1}{2}\gamma^{kl} \left(\partial_i \gamma_{jl} + \partial_j \gamma_{li} - \partial_l \gamma_{ij}\right).
\end{equation}

We use our analysis code \texttt{mescaline}, written to analyse three-dimensional HDF5 data output from our simulations. The code reads in the spatial metric $\gamma_{ij}$, the lapse $\alpha$, the extrinsic curvature $K_{ij}$, the density $\rho$, and the velocity $v^{i}$ from the Einstein Toolkit three-dimensional output. From these quantities we calculate the spatial Ricci tensor $R_{ij}$ from the spatial metric, and hence the Ricci scalar via $R=\gamma^{ij}R_{ij}$. We take the trace of the extrinsic curvature $K=\gamma^{ij}K_{ij}$ and with the set of equations defined in Appendix~\ref{sec:appx_avg} we calculate averages and the resulting backreaction terms. We also use \texttt{mescaline} to calculate the Hamiltonian and momentum constraint violation, discussed in Appendix~\ref{appx:constraints}. We compute derivatives using centred finite difference operators, giving second order accuracy in both space and time, the same order as the Einstein Toolkit's spatial discretisation.

\section{Results} \label{sec:results}
Figure~\ref{fig:256evol} shows time evolution of a two-dimensional slice of the density $\rho$ through the midplane of the $L=1$ Gpc domain at $256^{3}$ resolution. We show the growth of structures from $z=1100$ (top left) through to $z=0$ (bottom right). The $1\sigma$ variance in $\delta$ evolves from $\sigma_{\delta}=0.0026$ (top left) to $\sigma_{\delta}=3.92$ (bottom right). 

Figure~\ref{fig:128allsim_z0} shows two-dimensional slices through the midplane of three $128^{3}$ resolution simulations with domain size $L=1$ Gpc, 500 Mpc, and 100 Mpc (left to right), at redshift $z=0$. As we sample smaller scales we see a more prominent web structure forming. Our fluid treatment of dark matter implies over-dense regions continue to collapse towards infinite density, rather than forming virialised structures. This should, in general, yield a higher density contrast on small scales than we expect in the Universe.

Figure~\ref{fig:GRstuff} shows (left to right) the matter expansion rate $\theta$, the spatial Ricci curvature $R$, and the shear $\sigma^{2}$, respectively, at $z=0$. Each quantity is normalised to the global Hubble expansion $\mathcal{H}_{\rm all}$. The curvature and shear panels are normalised to correspond to the respective density parameters: $\Omega_R$ defined in \eqref{eqs:cosmic_qrtet}, and $\Omega_\sigma = \langle \sigma^2 \rangle / (3 \mathcal{H}_{\rm all}^2)$ defined in \citet{montanari2017}. We calculate $\theta$ using \eqref{eq:theta}, $\sigma^2$ using \eqref{eq:sigma_def} and \eqref{eq:sigma2}, and $R$ using the definitions \eqref{eq:ricci4Ddef} and \eqref{eq:christoffeldef}. Each panel shows a two-dimensional slice through the midpoint of the $L=1$ Gpc domain at $256^{3}$ resolution. Our relativistic quantities can be seen to closely correlate with the density distribution at the same time, shown in the bottom right panel of Figure~\ref{fig:256evol}.

\subsection{Global averages}
\begin{figure}
	\includegraphics[width=0.5\textwidth]{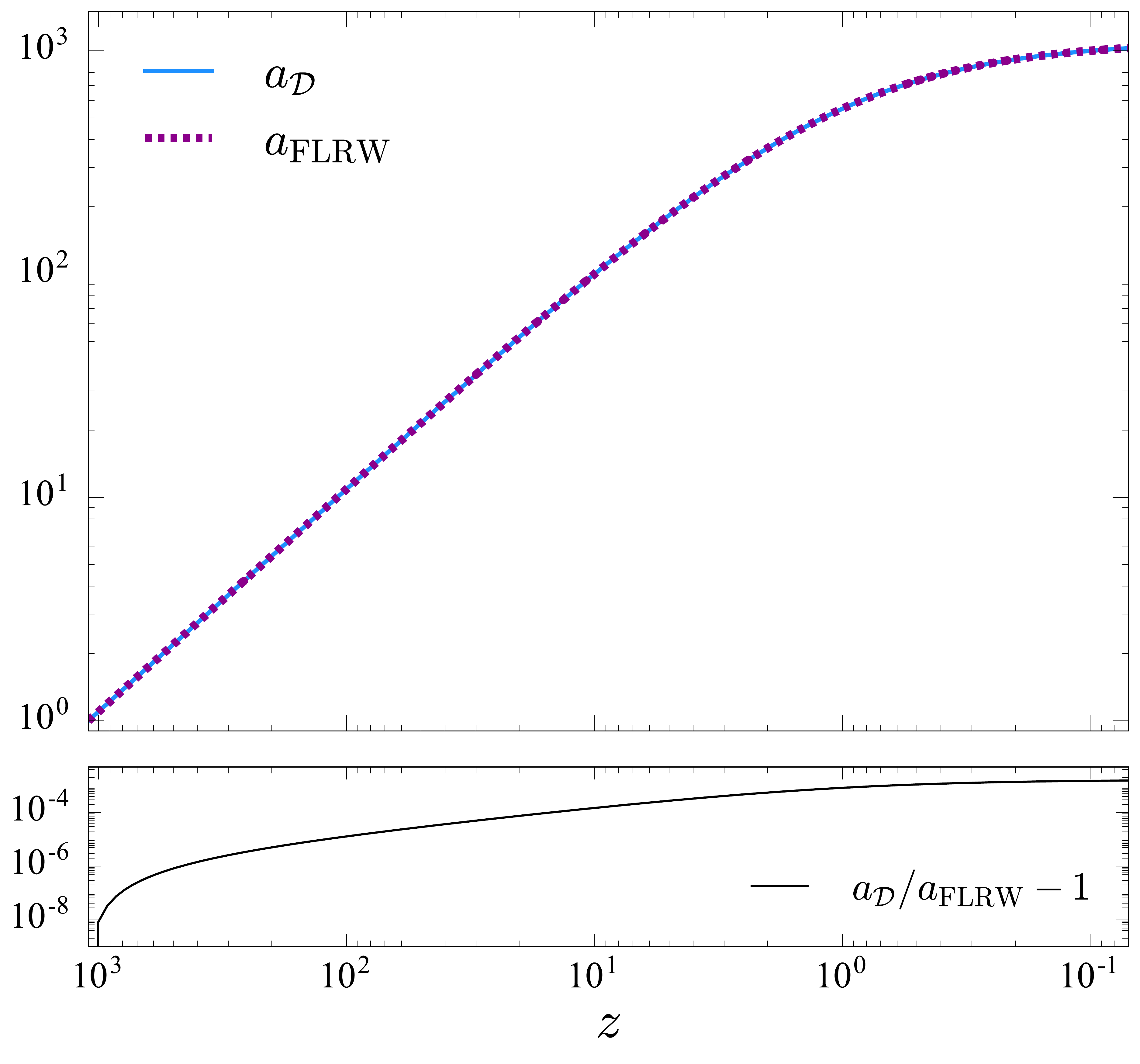}
    \caption{Globally, our expansion coincides with that of FLRW. The blue curve in the top panel shows the effective scale factor $a_\mathcal{D}$, calculated over the entire $L=1$ Gpc domain. The dashed magenta curve shows the equivalent FLRW solution (with $\Omega_m=1$), as a function of redshift. The bottom panel shows the residual error for this $256^3$ resolution calculation.}
    \label{fig:aDaFLRW}
\end{figure}

Figure~\ref{fig:aDaFLRW} shows the global evolution of the effective scale factor, $a_\mathcal{D}$. The blue curve shows $a_\mathcal{D}$ calculated over the whole $L=1$ Gpc, $256^{3}$ resolution domain with \eqref{eq:aDaDV}. The purple dashed curve in the top panel shows the corresponding FLRW solution for the scale factor, $a_\mathrm{FLRW}$, found by solving the Hamiltonian constraint for a flat, matter-dominated, homogeneous, isotropic Universe in the longitudinal gauge, 
\begin{equation}
	\frac{a'}{a} = \sqrt{\frac{8\pi G\bar{\rho}\,a^2}{3}},
\end{equation}
giving the solution \eqref{eq:confa}. The bottom panel of Figure~\ref{fig:aDaFLRW} shows the residual error between the two solutions, which remains below $10^{-3}$ for the evolution to $z=0$.

Analysing the cosmological parameters as an average over the entire simulation domain we find agreement with the corresponding FLRW model in our chosen gauge. Globally, at $z=0$, we find $\Omega_m \approx 1$, $\Omega_R \approx 10^{-8}$, and $\Omega_Q + \Omega_L \approx 10^{-9}$. Systematic errors on these values are discussed in Appendix~\ref{appx:convergence}.

\begin{figure*}
	\includegraphics[width=\textwidth]{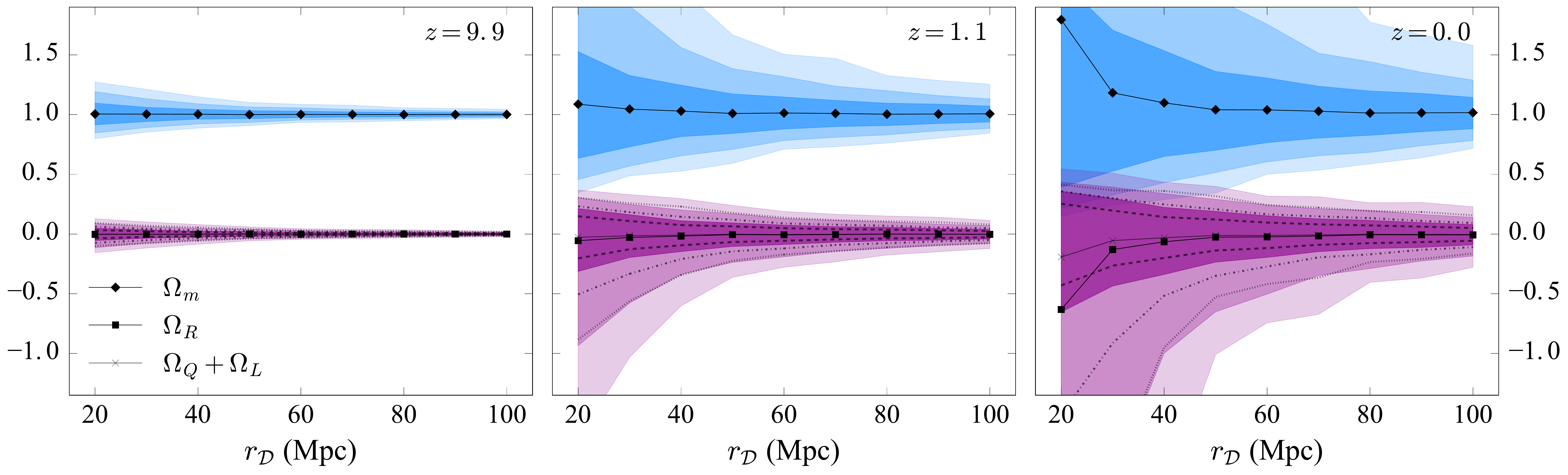}
    \caption{Growing inhomogeneity in matter, curvature, and backreaction. Here we show the cosmological parameters for spheres with various radii $r_{\mathcal{D}}$, randomly placed within an $L=1$ Gpc domain at $256^{3}$ resolution. Black points show mean values over 1000 spheres at each radius, progressively lighter blue and purple shaded regions show the 68\%, 95\%, and 99.7\% confidence intervals for $\Omega_{m}$ and $\Omega_{R}$, respectively. Crosses show the mean contribution from backreaction terms $\Omega_{Q}+\Omega_{L}$, while dashed, dot-dashed, and dotted lines show the 68\%, 95\%, and 99.7\% confidence intervals, respectively. Left to right panels are redshifts $z = 9.9,1.1$, and $0.0$, respectively.}
    \label{fig:omegas_big}
\end{figure*}
\begin{figure}
	\includegraphics[width=0.5\textwidth]{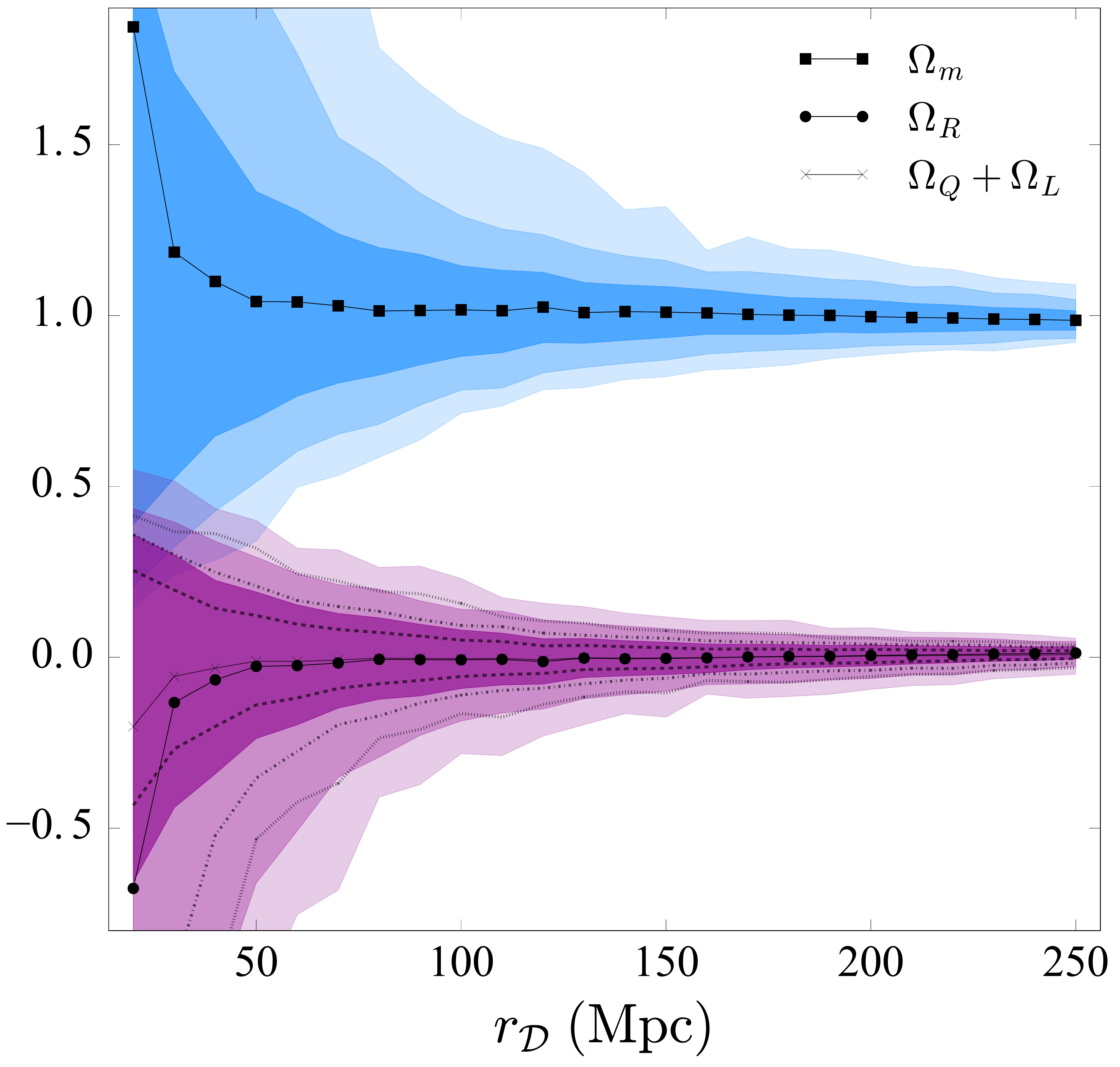}
    \caption{We approach homogeneity when averaging over larger scales. Here we show the right-most panel of Figure~\ref{fig:omegas_big} extending to averaging radius $r_\mathcal{D}=250$ Mpc. Black points show the mean $\Omega_m$, $\Omega_R$, and $\Omega_Q+\Omega_L$ over the 1000 spheres at each radius. Progressively lighter blue and purple shaded regions show the 68\%, 95\%, and 99.7\% confidence intervals for $\Omega_m$ and $\Omega_R$, while dashed, dot-dashed, and dotted lines show these for $\Omega_Q+\Omega_L$.}
    \label{fig:omegas_larger}
\end{figure}
\subsection{Local averages}
\subsubsection{Cosmological parameters}
Figure~\ref{fig:omegas_big} shows cosmological parameters calculated within spheres of various averaging radii, $r_{\mathcal{D}}$, within an $L=1$ Gpc domain at $256^{3}$ resolution. Left to right panels correspond to increasing time (decreasing $z$), showing $z=9.9,1.1$, and $0$, respectively. Black points show the mean value over 1000 spheres at the corresponding averaging radius, showing filled circles for $\Omega_{m}$, filled squares for $\Omega_{R}$, and crosses for $\Omega_{Q} + \Omega_{L}$. Over these 1000 spheres we also show the 68\%, 95\%, and 99.7\% confidence intervals for $\Omega_{m}$ and $\Omega_{R}$ as progressively lighter blue and purple shaded regions, respectively. The same confidence intervals for the contribution from the backreaction terms, $\Omega_{m} + \Omega_{L}$, are shown as dashed, dot-dashed, and dotted lines respectively.  Figure~\ref{fig:omegas_larger} shows the same calculation of the cosmological parameters at $z=0$, extending averaging radii to $r_\mathcal{D}=250$ Mpc. 

At redshift $z=0$, considering averaging radii corresponding to the approximate homogeneity scale of the Universe \citep{scrimgeour2012}, $80<r_\mathcal{D}<100\,h^{-1}$Mpc, we find $\Omega_{m}=1.01\pm0.09 $, $\Omega_{R}=-0.006\pm0.06$, and $\Omega_{Q}+\Omega_{L}=-0.004\pm0.04$. These are the mean values over all spheres with $r_\mathcal{D}=80-100\,h^{-1}$Mpc; 3000 spheres in total. Variations are the 68\% confidence intervals of the distribution.

Below the measured homogeneity scale, with $r_\mathcal{D}<100\,h^{-1}$Mpc, we use 13 individual radii each with a sample of 1000 spheres. We find $\Omega_m=1.1^{+0.12}_{-0.31}$, $\Omega_R=-0.08^{+0.21}_{-0.06}$, and $\Omega_{Q}+\Omega_{L}=-0.03^{+0.11}_{-0.06}$.

Considering scales above this homogeneity scale, we use $100<r_\mathcal{D}<180\,h^{-1}$Mpc with a total of 11 radii sampled and 1000 spheres each. On these scales we find $\Omega_m=0.997\pm0.05$, $\Omega_R=0.005\pm0.03$, and $\Omega_{Q}+\Omega_{L}=0.003\pm0.02$.

Systematic errors in all quoted cosmological parameters here are discussed in Appendix~\ref{appx:convergence}.

\begin{figure}
	\includegraphics[width=0.5\textwidth]{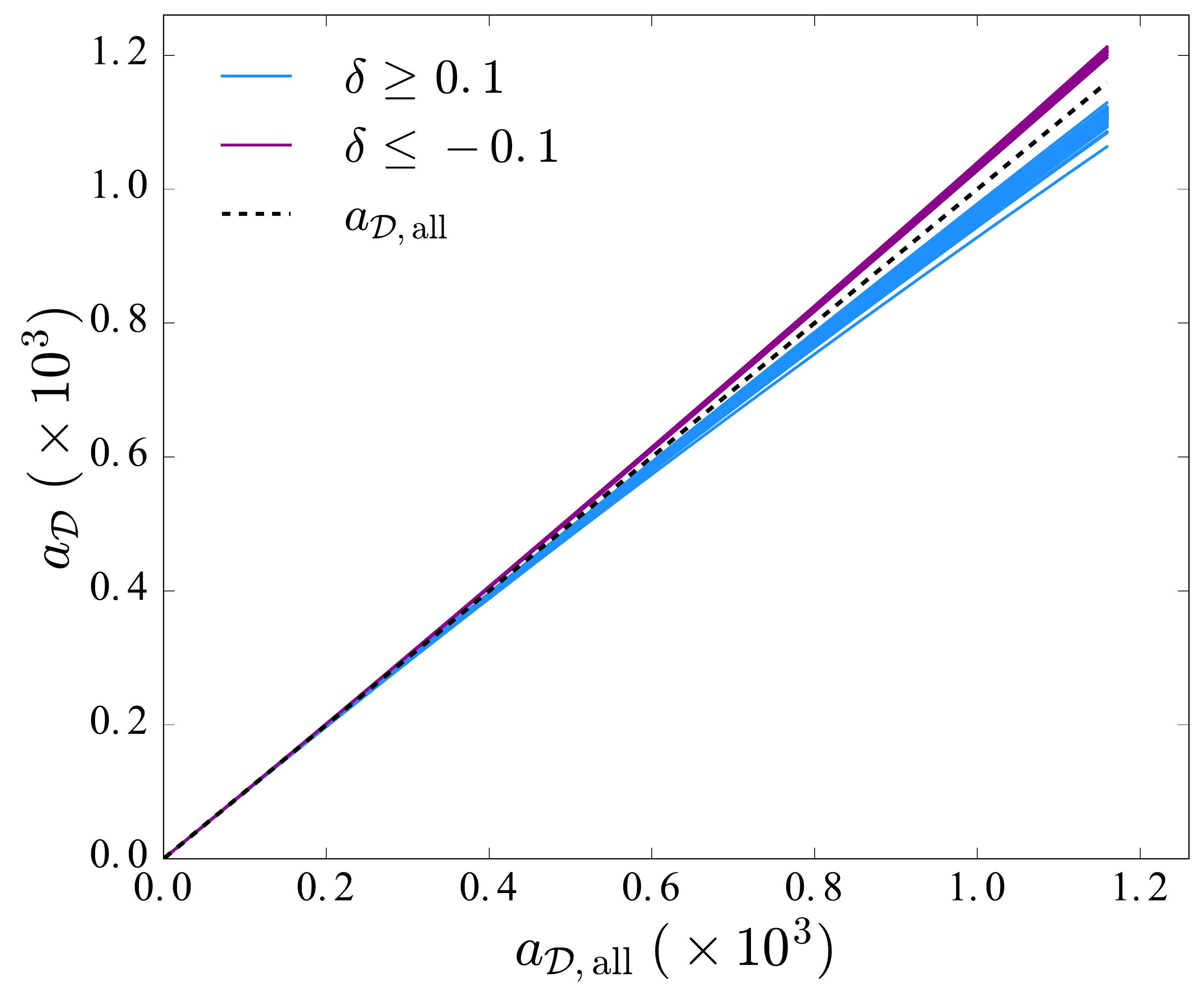}
    \caption{Inhomogeneous expansion as a function of time, showing the effective scale factor $a_\mathcal{D}$ calculated in spheres of radius 100 Mpc as a function of global expansion $a_{\mathcal{D},\mathrm{all}}$. We calculate $a_\mathcal{D}$ in an $L=500$ Mpc simulation at $128^{3}$ resolution. Blue curves show overdense regions with $\delta\geq0.1$, while purple curves show underdense regions with $\delta\leq-0.1$. The black dashed line shows the mean expansion over the whole domain.}
    \label{fig:aDr100}
\end{figure}
\begin{figure*}
	\includegraphics[width=\textwidth]{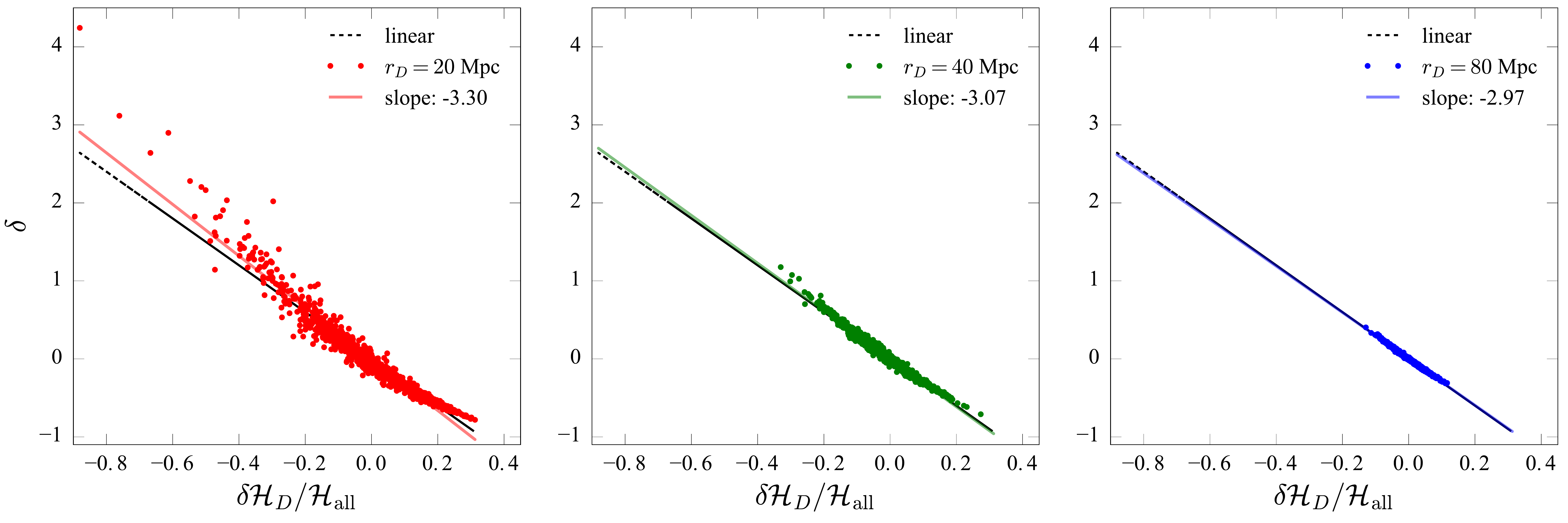}
    \caption{Relation between the fractional density perturbation, $\delta$, and the deviation in the Hubble parameter, $\delta \mathcal{H_D}/\mathcal{H}_\mathrm{all}$, for averaging radii $r_\mathcal{D}=20,40$, and 80 Mpc (left to right), respectively. Points in each panel represent individual spheres of 1000 sampled at each radius, and the solid line of the same colour is the best-fit linear relation, with slope indicated in each panel. The black dashed line is the prediction from linear theory.}
    \label{fig:dHvsdelta}
\end{figure*}

\subsubsection{Scale factor}
Figure~\ref{fig:aDr100} shows the evolution of the effective scale factor calculated within spheres of $r_\mathcal{D}=100$ Mpc, relative to the global value $a_{\mathcal{D},\mathrm{all}}$, which we use as a proxy for time. The dashed line shows the global average, blue curves show $a_\mathcal{D}$ for overdense regions with $\delta\geq0.1$, and purple curves for underdense regions with $\delta\leq-0.1$. In total, we sample 1000 spheres with randomly placed (fixed) origins within an $L=1$ Gpc, $256^{3}$ resolution simulation. Underdense regions with $\delta\leq-0.1$ expand $4-5$\% faster than the mean at $z=0$, while overdense regions with $\delta\geq0.1$ expand $2-8$\% slower.

\subsubsection{Hubble parameter}
Figure~\ref{fig:dHvsdelta} shows the relation between the density, $\delta$, of a spherical domain and the corresponding deviation in the Hubble parameter $\delta H_\mathcal{D}/\bar{H}_\mathcal{D}=(H_\mathcal{D}-\bar{H}_\mathcal{D})/\bar{H}_\mathcal{D}$; the expansion rate of that sphere. We show the density and variation in the Hubble parameter for averaging radii $r_\mathcal{D}=20,40$, and 80 Mpc, (left to right) respectively. Points in each panel show individual measurements within 1000 randomly placed spheres of the same radius. The solid line of the same colour in each panel is the linear best-fit for the data points, with slope indicated in each panel. 

Linear perturbation theory predicts the relation between the average density, $\langle\delta\rangle$, of a spherical perturbation and the deviation from the Hubble flow of that spherical region, $\delta \mathcal{H_D}/\mathcal{H}_\mathrm{all}$, to be \citep{lahav1991}
\begin{equation} \label{eq:lindHvsdelta}
	\langle\delta\rangle = -3\,F\,\frac{\delta \mathcal{H_D}}{\mathcal{H}_\mathrm{all}},
\end{equation}
where $F=\Omega_m^{0.55}$ is the growth rate of matter \citep{linder2005}, which for our global average $\Omega_m\approx1$ is $F=1$. This in turn implies that the growth rate of structures in our simulations is larger than in the $\Lambda$CDM Universe where $\Omega_m\approx0.3$ \citep[e.g.][]{DESCollab2017a,bonvin2017,planck2016params,bennett2013}. The black dashed line in each panel of Figure~\ref{fig:dHvsdelta} is the relation \eqref{eq:lindHvsdelta}, a slope of -3. On 20 Mpc scales the line of best fit is 10\% larger than this prediction, on 40 Mpc scales it is 2.2\% larger, and on 80 Mpc scales is 0.9\% smaller.

\section{Discussion}\label{sec:discussion}
We have presented simulations of nonlinear structure formation with numerical relativity, beginning with initial conditions drawn from the CMB matter power spectrum. These simulations allow us to analyse the effects of large density contrasts on the surrounding spacetime, and consequently on cosmological parameters. We calculate the cosmological parameters $\Omega_m$, $\Omega_R$, $\Omega_Q,$ and $\Omega_L$, together describing the content of the Universe, for spherical subdomains embedded within a 256$^{3}$ resolution, $L=1$ Gpc simulation. We vary the averaging radius between $20\leq r_\mathcal{D}\leq250$ Mpc, representing scales both below and above the measured homogeneity scale of the Universe.

Our results were obtained using simulations sampling the matter power spectrum down to scales of two grid points. Quantifying the errors in such a calculation is difficult because structure formation occurs fastest on small scales, implying different physical structures at different resolutions. This is a known problem in cosmological simulations, not unique to General Relativistic cosmology \citep[see e.g.][]{schneider2016}. To correctly quantify such errors, we must maintain the same density gradients between several simulations at different computational resolution. This becomes difficult when the perturbations themselves are nonlinear. Even with identical initial conditions, we see a different distribution of structures at redshift $z=0$ when sampling nonlinear scales at different resolutions. To approximate the errors on our main results, we instead analyse a set of test simulations in which we simulate a fixed amount of large-scale structure (see Appendix~\ref{appx:convergence}). This allows for a reliable Richardson extrapolation of the solution to approximate the error in our main results at redshift $z=0$.



Regardless of this, the main result of this paper is that we find $\Omega_m\approx1$ in all simulations we analyse here. Any unquantified errors are unlikely to significantly shift this result,  and all global effects of backreaction and curvature are likely to remain small with an improved sampling of small scales.

\subsection{Global averages}
We find global cosmological parameters consistent with a matter-dominated, flat, homogeneous, isotropic universe, and therefore no global backreaction. 
The evolution of the effective scale factor $a_\mathcal{D}$, evaluated over the whole domain, coincides with the corresponding FLRW model, as shown in Figure~\ref{fig:aDaFLRW}. The $<10^{-3}$ discrepancy between the two solutions does not correlate with the onset of nonlinear structure formation, indicating that this difference is most likely computational error. 

We find a globally flat geometry in our simulations with $\Omega_R \approx 10^{-8}$. This could be a result of our treatment of the matter as a fluid. We cannot create virialised objects and so any "clusters" will continue to collapse towards infinite density. In reality, a dark matter halo or galaxy cluster would form, be supported by velocity dispersion, and stop collapsing. The surrounding voids would continue to expand and potentially contribute to a globally negative curvature \citep[see e.g.][]{bolejko2017b,bolejko2018a}. Without a particle description for dark matter alongside numerical relativity we cannot properly capture this effect.

Any contribution from backreaction, $\mathcal{Q_D}$ or $\mathcal{L_D}$, is due to variance in the expansion rate and shear. The left panel of Figure~\ref{fig:GRstuff} shows the matter expansion rate $\theta$, where collapsing regions (yellow, orange, and red) balance the expanding regions (green) due to our treatment of matter. While we see spatial variance in $\theta$, there is no global contribution from backreaction under our assumptions. Therefore, in our chosen gauge and under the caveats described in Section~\ref{sec:caveat} below, backreaction from structure formation is unlikely to explain dark energy. 


\subsection{Local averages}
We find strong positive curvature on scales below the homogeneity scale of the Universe. Variations in measured cosmological parameters are up to 31\% based purely on location in an inhomogeneous matter distribution. Our result is similar to that of \citet{bolejko2017b} on small scales, but with larger variance in $\Omega_R$ because of increased small-scale density fluctuations due to our fluid treatment of dark matter.

On the approximate homogeneity scale of the Universe we find mean cosmological parameters consistent with the corresponding FLRW model to $\sim1\%$. Aside from this, we find the parameters can deviate from these mean values by 4-9\% depending on physical location in the simulation domain. This implies that, although on average these coincide with a flat, homogeneous, isotropic Universe, an observers interpretation may differ by up to 9\% based purely on her position in space. 

As we approach larger averaging radii within a 1 Gpc$^3$ volume, we begin to move away from independent spheres, and each sphere begins to overlap with others; effectively sampling the same volume. Due to this, the confidence intervals contract, and eventually at $r_\mathcal{D}\approx400$ Mpc most spheres become indistinguishable from the mean. The beginning of this is evident in Figure~\ref{fig:omegas_larger} as we approach $r_\mathcal{D}=250$ Mpc. This transition appears to be due to overlapping spheres, although could in part be due to the statistical homogeneity of the matter distribution at these scales.

Local observations of type 1a supernovae generally probe scales of $75-450\,h^{-1}$Mpc \citep{wuhuterer2017}. Nearby objects are excluded from the data in an effort to minimise cosmic variance on the result \citep{riess2016,riess2018a,riess2018b}. In this work, we cannot meaningfully sample scales above 250 Mpc because our maximum domain size is only 1 Gpc$^3$. In order to sample all scales used in nearby SNe surveys, we would need a domain size of $L\gtrsim 10\,h^{-1}$Gpc, with a resolution up to $1024^{3}$. Current computational constraints, and the overhead of numerical relativity, currently restrict us to domain sizes and resolutions used in this work. To address scales as similar as possible to those used in local surveys, we consider $75<r_\mathcal{D}<180\,h^{-1}$Mpc. On these scales we find $\Omega_m=1.002\pm0.06$, $\Omega_R=0.002\pm0.04$, and $\Omega_{Q}+\Omega_{L}=0.001\pm0.02$, where variances are the 68\% confidence intervals due to local inhomogeneity. This implies based on an observers physical location, measured deviations from homogeneity on these scales could be up to 6\%. We expect this variance to decrease when including the full range of observations; including radii up to $450\,h^{-1}$Mpc. We investigate this further, including extrapolation to larger scales, in our companion paper \citet{macpherson2018b}.

While the global effective scale factor demonstrates pure FLRW evolution, we find inhomogeneous expansion within spheres of 100 Mpc radius. Figure~\ref{fig:aDr100} shows the expansion rate differs by $2-8\%$ depending on the relative density of the region sampled. These differences agree with linear perturbation theory, to within 1\%, on $\gtrsim80$ Mpc scales, with smaller scales showing differences of up to 10\%. These differences are most likely due to the nonlinearity of the density field on these scales, although, in addition, could involve general relativistic corrections. To properly test this we would require an equivalent Newtonian cosmological simulation to compare this relation at nonlinear scales, which we leave to future work.

\subsection{Caveats}\label{sec:caveat}
\begin{enumerate}
	\item Our treatment of dark matter as a fluid is the main limitation of this work. Under this assumption, we are unable to form bound structures supported from collapse by velocity dispersions. In cosmological N-body simulations, particle methods are adopted so as to capture the formation of galaxy haloes, and local groups of galaxies as bound structures. Adopting a fully general relativistic framework in addition to particle methods would allow us to adopt a proper treatment of dark matter in parallel with inhomogeneous expansion. 
	\item We take averages over purely spatial volumes. In reality, an observer would measure her past light cone, and hence the evolving Universe. Our results can thus be considered an upper limit on the variance due to inhomogeneities, since any structures located in the past light cone will be more smoothed out.
	\item Our results are explicitly dependent on the chosen averaging hypersurface. The result of averaging across different hypersurfaces has been investigated \citep{adamek2017,giblin2018}, and the results can show significant differences. It is clear the physical choice of hypersurface can be important for quantifying the backreaction effect.
	\item We assume $\Lambda=0$, and begin our simulations assuming a flat, matter dominated background cosmology with small perturbations. Throughout the evolution, on a global scale, we find the average $\Omega_m\approx1$; consistent with this model. It is extremely well constrained that our Universe is best described by a matter content $\Omega_m\approx0.3$ \citep[e.g.][]{DESCollab2017a,bonvin2017,planck2016params,bennett2013}. The growth rate of cosmological structures in our simulations will therefore be amplified relative to the $\Lambda$CDM Universe. 
	\item Given our limited spatial resolution, we underestimate the amount of structure compared to the real Universe. In addition, we resolve structures down to scales of two grid points, which means these structures may be under resolved.
\end{enumerate}

\section{Conclusions} \label{sec:conclude}
We summarise our findings as follows:
\begin{enumerate}
	\item We find no global backreaction under our assumptions. Over the entire simulation domain we have $\Omega_m \approx 1$, $\Omega_R \approx 10^{-8}$, and $\Omega_Q + \Omega_L \approx 10^{-9}$, in our chosen gauge; consistent with a matter-dominated, flat, homogeneous, isotropic universe.
	\item We find strong deviation from homogeneity and isotropy on small scales. Below the measured homogeneity scale of the Universe ($r_\mathcal{D}\lesssim 100\,h^{-1}$Mpc) we find deviations in cosmological parameters of $6-31\%$ based purely on an observers physical location.
	\item Above the homogeneity scale of the universe ($100<r_\mathcal{D}<180\,h^{-1}$Mpc) we find mean cosmological parameters coincide with the corresponding FLRW model, with potential $2-5\%$ deviations due to inhomogeneity.
	\item We find agreement with linear perturbation theory within 1\% on $\geq80$ Mpc scales for the relation between the density of a spherical region and its corresponding deviation from the Hubble flow. However, these few percent deviations on smaller scales may prove important in forthcoming cosmological surveys.
\end{enumerate}
While we find no global backreaction in our cosmological simulations, our numerical relativity calculations show significant contributions from curvature and other nonlinear effects on small scales. 

\begin{acknowledgments}
We thank the anonymous referee for their comments that significantly improved the quality of this manuscript. We thank Chris Blake, Marco Bruni, Krzysztof Bolejko, Syksy R\"{a}s\"{a}nen, David Wiltshire, Eloisa Bentivegna, Chris Clarkson, Ruth Durrer, Timothy Clifton, Jim Mertens, and Tom Giblin for useful feedback and discussions, in general, as well as specific to this work. HM especially thanks Marco Bruni and the University of Portsmouth for financial support and hospitality during the production of this work. HM thanks the organisers and participants of the Inhomogeneous Cosmologies conference in Toru\'n 2017 for their feedback and support. We use the Riemannian Geometry \& Tensor Calculus (RGTC) package for Mathematica, written by Sotirios Bonanos. This work was supported by resources provided by the Pawsey Supercomputing Centre with funding from the Australian Government and the Government of Western Australia. HM thanks the Astronomical Society of Australia for their funding support that helped contribute to this work. PDL is supported through Australian Research Council (ARC) Future Fellowship FT160100112 and ARC Discovery Project DP180103155.  DJP is supported through ARC FT130100034. 
\end{acknowledgments}

\appendix

%

\section{Averaging in the non-comoving gauge} \label{sec:appx_avg}
Averaging Einstein's equations in a non-comoving gauge results in the averaged Hamiltonian constraint
\begin{equation}
	6\mathcal{H_D}^{2} = 16\pi\langle\Gamma^{4}\rho\rangle - \mathcal{R_D} - Q_\mathcal{D} + \mathcal{L_D},
\end{equation}
where we define
\begin{align}
	\mathcal{R_D} &\equiv \langle\Gamma^{2}R\rangle, \label{eq:RD}\\
	Q_\mathcal{D} &\equiv \frac{2}{3} \left( \langle\theta^{2}\rangle - \langle\theta\rangle^{2}\right) - 2\langle\sigma^{2}\rangle, \label{eq:QD}\\
	\mathcal{L_D} &\equiv 2\langle\sigma_{B}^{2}\rangle - \frac{2}{3}\langle\theta_{B}^{2}\rangle - \frac{4}{3}\langle\theta\theta_B\rangle. \label{eq:LD}
\end{align}
Here, $\Gamma=1/\sqrt{1-v^{i}v_{i}}$ is the Lorentz factor, $R\equiv \gamma^{ij}R_{ij}$ is the three-dimensional Ricci curvature of the averaging hypersurfaces, with $R_{ij}$ the spatial Ricci tensor. Here 
\begin{equation} \label{eq:sigma2}
	\sigma^{2}=\frac{1}{2}\sigma^{i}_{\phantom{i}j}\sigma^{j}_{\phantom{j}i},
\end{equation}
where $\sigma_{ij}$ is the shear tensor, defined as
\begin{equation} \label{eq:sigma_def}
	\sigma_{\mu\nu}\equiv h^{\alpha}_{\phantom{\alpha}\mu}h^{\beta}_{\phantom{\beta}\nu} \nabla_{(\alpha} u_{\beta)} - \frac{1}{3}\theta h_{\mu\nu}.
\end{equation}
As in \citep{umeh2011}, we introduce for simplification
\begin{subequations} \label{eqs:tensor_defs}
	\begin{align}
		\sigma_B^{2} &= \frac{1}{2}\sigma^{i}_{\phantom{i}Bj}\sigma^{j}_{\phantom{j}Bi} + \sigma_{ij}\sigma_{B}^{ij} \\
		\sigma_{Bij} &\equiv -\Gamma \beta_{ij} - \Gamma^{3}\left(B_{(ij)} - \frac{1}{3}Bh_{ij}\right) \\
		\beta_{\mu\nu} &\equiv h^{\alpha}_{\phantom{\alpha}\mu}h^{\beta}_{\phantom{\beta}\nu}\nabla_{(\alpha}v_{\beta)} - \frac{1}{3}\kappa h_{\mu\nu} \\
		B_{\mu\nu} &\equiv \frac{1}{3}\kappa(v_\mu n_\nu + v_\mu v_\nu) + \beta_{\alpha\mu}v^{\alpha}n_\nu + \beta_{\alpha\mu}v^{\alpha}v_\nu \\
				&+ W_{\alpha\mu}v^{\alpha}n_\nu + W_{\alpha\mu}v^{\alpha}v_\nu,		
	\end{align}
\end{subequations}
where we also define
\begin{subequations}\label{eqs:kapWB_defs}
    \begin{align} 
    	\kappa &\equiv h^{\alpha\beta}\nabla_{\alpha}v_{\beta}, \quad W_{\mu\nu} \equiv h^{\alpha}_{\phantom{\alpha}\mu}h^{\beta}_{\phantom{\beta}\nu}\nabla_{[\alpha}v_{\beta]}, \\
    	B &= \frac{1}{3}\kappa v^{\alpha}v_{\alpha} + \beta_{\mu\nu}v^{\mu}v^{\nu}.
    \end{align}
\end{subequations}
For a given tensor $A_{\mu\nu}$ we adopt the notation $A_{(\mu\nu)} = \frac{1}{2}(A_{\mu\nu} + A_{\nu\mu})$ and $A_{[\mu\nu]} = \frac{1}{2}(A_{\mu\nu} - A_{\nu\mu})$.

\section{Effective scale factors} \label{sec:appx_expn}
The effective expansion of an inhomogeneous domain can be defined by
\begin{align}
	\frac{\partial_\eta \,a_\mathcal{D}^{V}}{a_\mathcal{D}^{V}} &\equiv \frac{1}{3}\frac{\partial_\eta V_\mathcal{D}}{V_\mathcal{D}}, \\
	\Rightarrow a_\mathcal{D}^{V} &= \left( \frac{V_\mathcal{D}(\eta)}{V_\mathcal{D}(\eta_\mathrm{init})} \right)^{1/3},
\end{align}
where $V_\mathcal{D}(\eta)$ is the volume of the domain $\mathcal{D}$ at a given conformal time. The physical interpretation of this scale factor depends on the chosen hypersurface of averaging. If we choose the averaging surface to be comoving with the fluid; a surface with normal $u^{\mu}$, then the scale factor $a_\mathcal{D}^{V}$ describes the effective expansion of the fluid averaged over the domain. We define the averaging surface to be comoving with a set of observers with normal $n^{\mu}$; \textit{not} coinciding with $u^{\mu}$. In this case, $a_\mathcal{D}^{V}$ describes the expansion of the volume element, not of the fluid itself.

We define the Hubble parameter as the expansion of the fluid projected into the gravitational rest frame; the frame of observers with normal $n^{\mu}$. From this we define the effective scale factor of the fluid, $a_\mathcal{D}$ in \eqref{eq:aDdef}. We can relate the two scale factors by first considering the rate of change of the volume (with $\beta^{i}=0$) in the (3+1) formalism \citep{larena2009b},
\begin{equation}
	\frac{\partial_\eta V_{\mathcal{D}}}{V_{\mathcal{D}}} = \langle \,\alpha\;\Gamma^{-1}\;(\theta - \kappa)\,\rangle.
\end{equation}
Now, with $\partial_\eta a_\mathcal{D}/a_\mathcal{D} = \partial_\eta \mathrm{ln}(a_\mathcal{D})$, we can write
\begin{align}
	\partial_\eta \mathrm{ln}(a_{\mathcal{D}}) &= \frac{1}{3}\langle\theta\rangle, \label{eq:aDlog} \\
	\partial_\eta \mathrm{ln}(a_{\mathcal{D}}^{V}) &= \frac{1}{3}\langle \,\alpha\;\Gamma^{-1}\;(\theta - \kappa)\,\rangle, \label{eq:aDVlog}
\end{align}
subtracting \eqref{eq:aDVlog} from \eqref{eq:aDlog} we arrive at the relation
\begin{equation} \label{eq:aDfromaDV}
	a_{\mathcal{D}} = a_{\mathcal{D}}^{V} \;\mathrm{exp}\left(-\frac{1}{3}\int \langle \,\alpha\;\Gamma^{-1}\;(\theta - \kappa) - \alpha\theta\,\rangle \;{\rm d}\eta\right).
\end{equation}
Here, $a_{\mathcal{D}}^{V}$ is found by calculating the volume of the domain relative to the initial volume. Figure~\ref{fig:aDaFLRW} shows the evolution of \eqref{eq:aDfromaDV} (blue solid curve) as a function of redshift for a $256^{3}$ simulation over an $L=1$ Gpc domain, relative to the equivalent FLRW solution (purple dashed curve).

\begin{centering}
\begin{figure*}
\begin{subfigure}{0.45\textwidth}
  \includegraphics[width=0.9\textwidth]{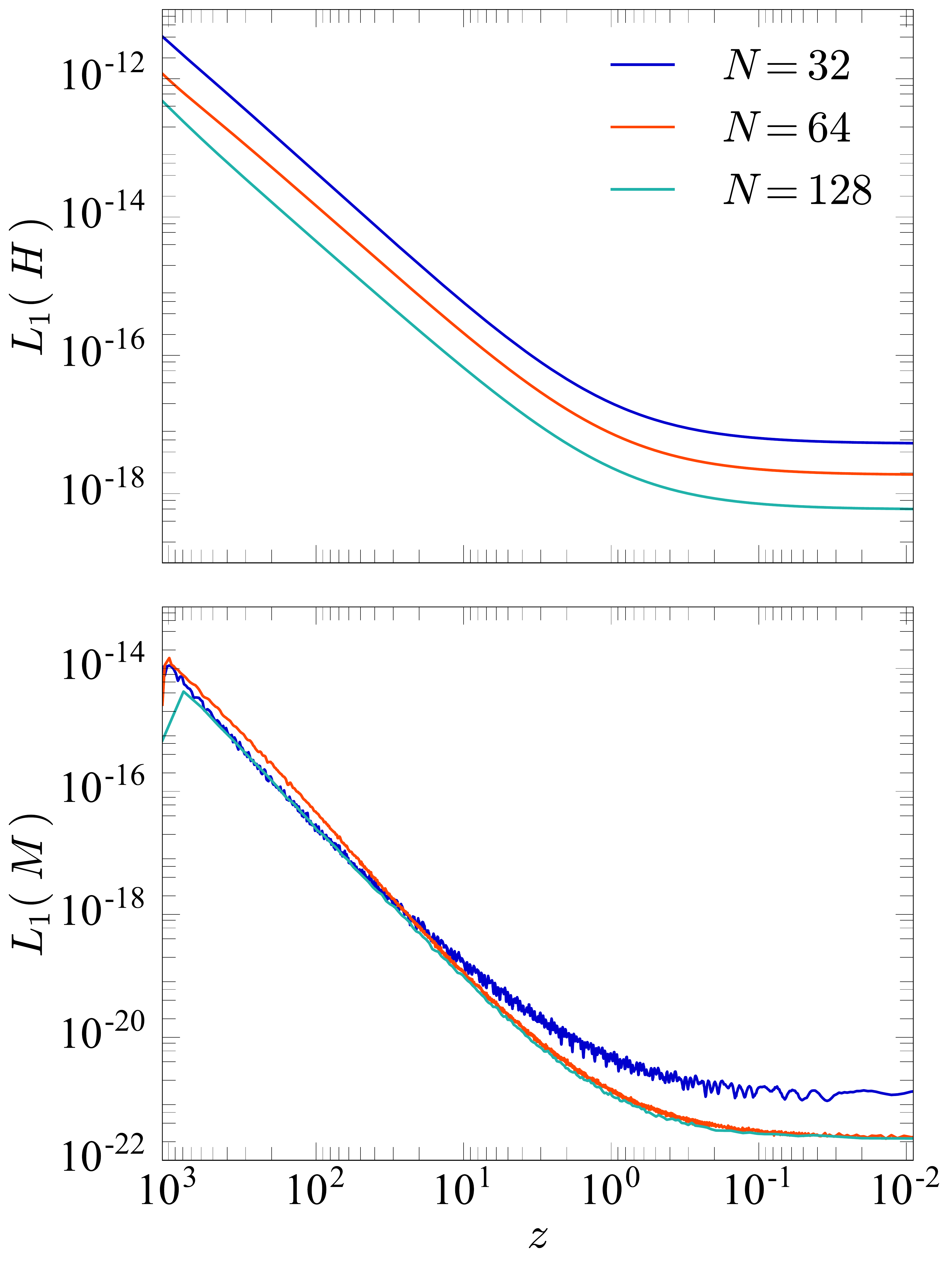}
  \label{fig:constraints_controlRaw}
\end{subfigure}%
\begin{subfigure}{0.45\textwidth}
  \includegraphics[width=0.9\textwidth]{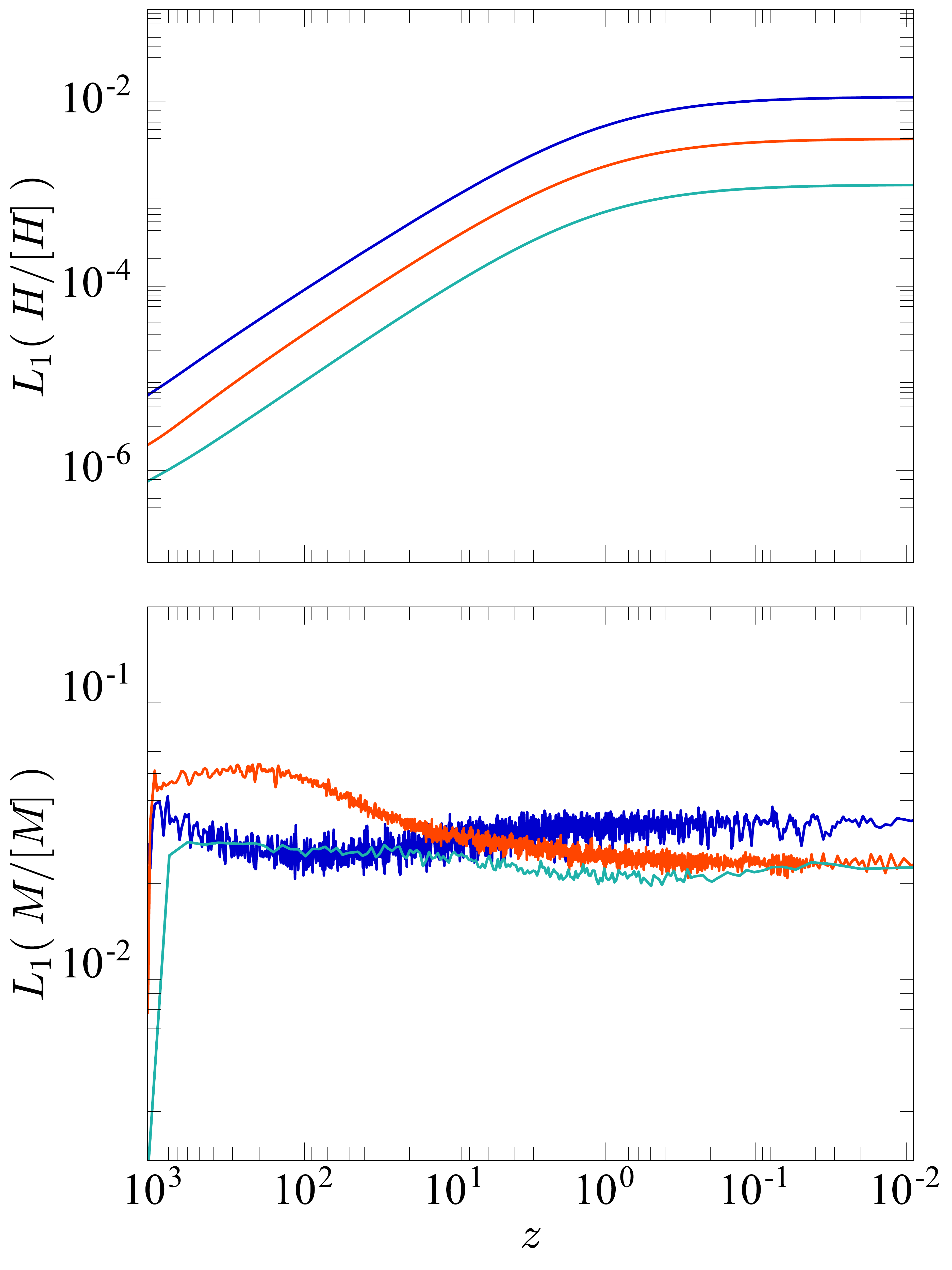}
   \label{fig:constraints_controlRel}
\end{subfigure}
\caption{Relative (right panels) and raw (left panels) constraint violation as a function of effective redshift calculated using \eqref{eq:l1Herror_rel} and \eqref{eq:l1Herror_raw}, respectively. We show violations for the simulations with a controlled number of physical modes. Top panels show the Hamiltonian constraint violation, and bottom panels show the momentum constraint violation. Colours show different resolutions as indicated by the legend.}
 \label{fig:constraints_control}
\end{figure*}
\end{centering}

\begin{centering}
\begin{figure}
\begin{subfigure}{0.5\textwidth}
     \includegraphics[width=0.9\textwidth]{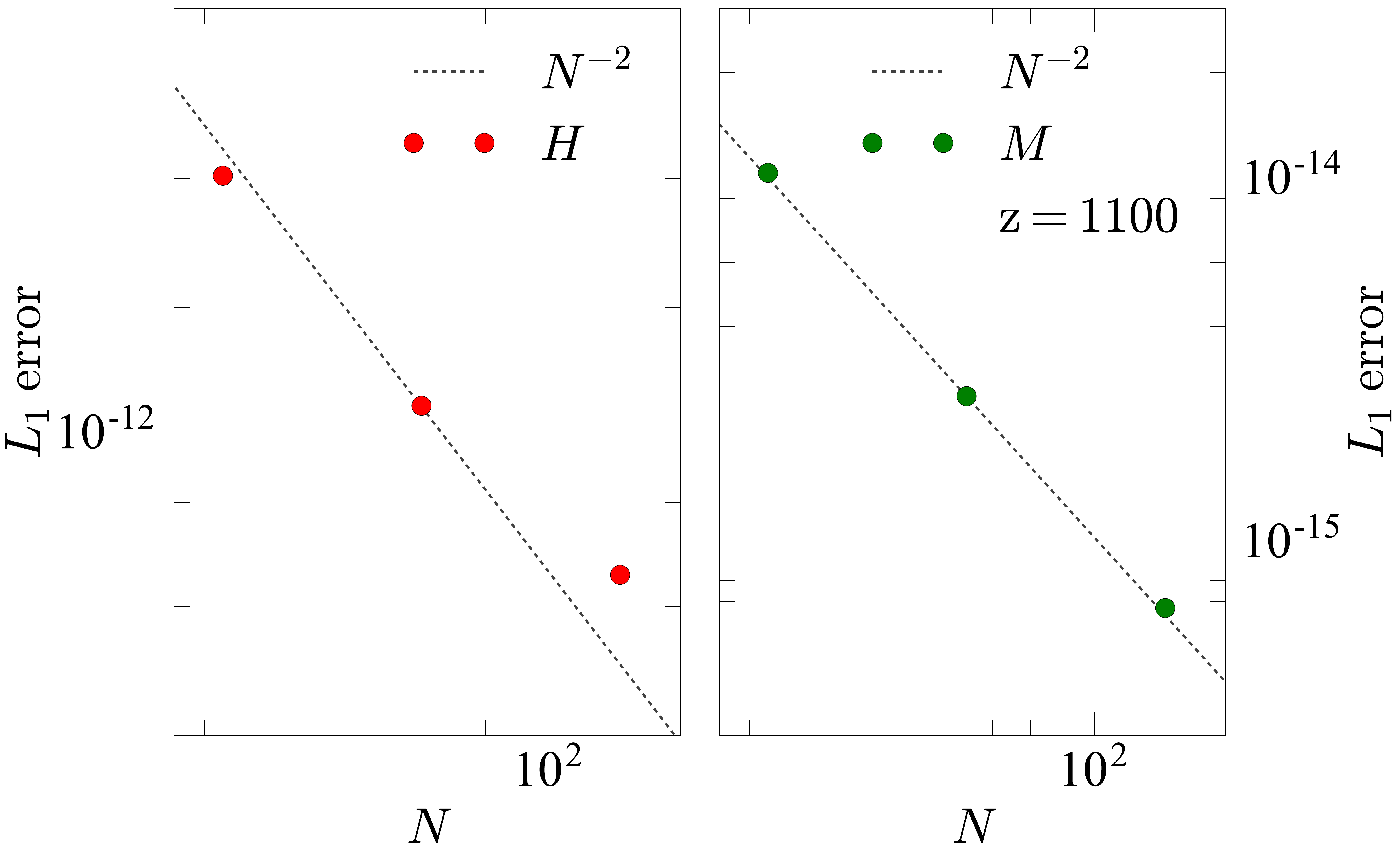}
\end{subfigure}

\begin{subfigure}{0.5\textwidth}
     \includegraphics[width=0.9\textwidth]{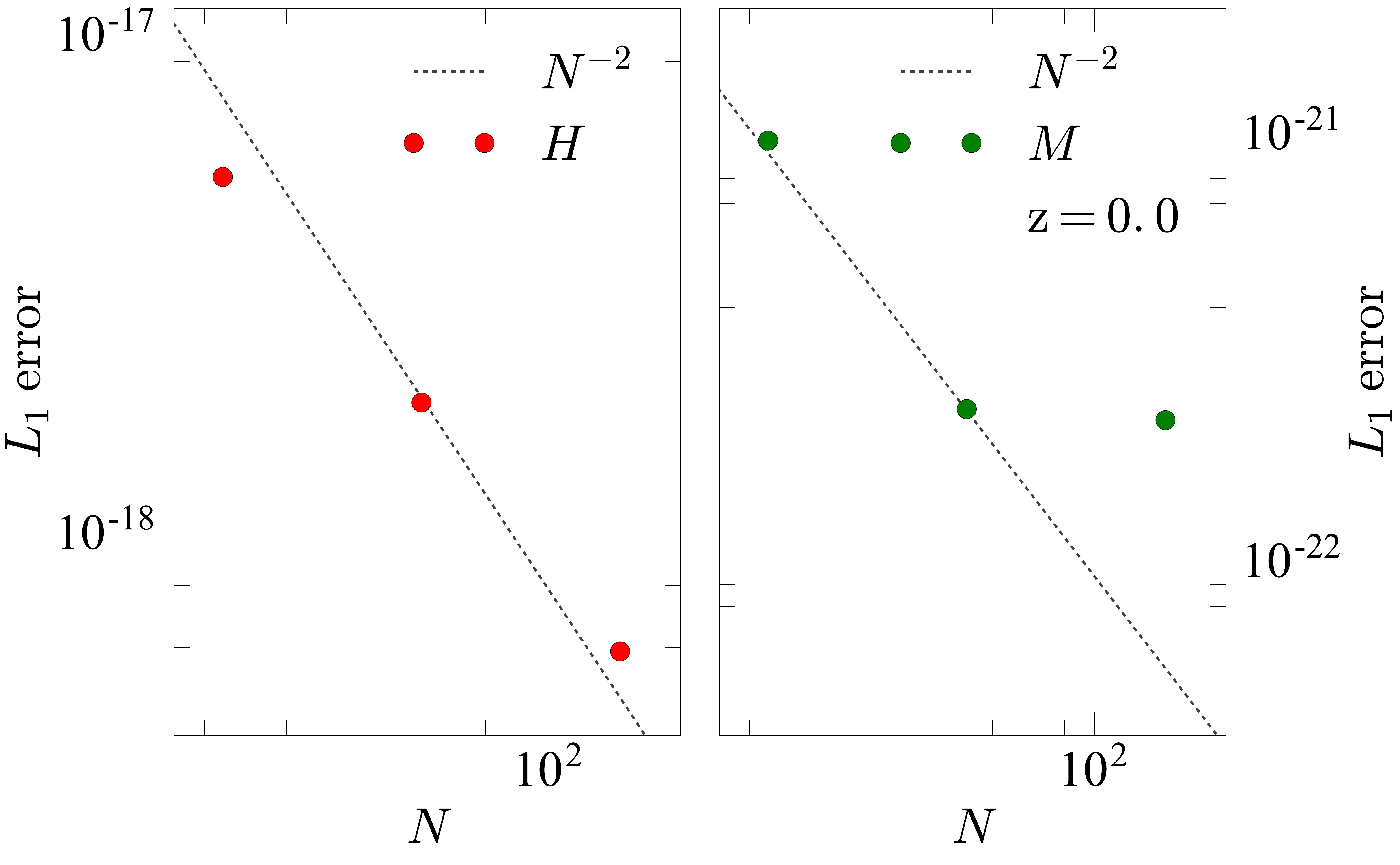}
\end{subfigure}
    \caption{Second order convergence of the Hamiltonian and momentum constraints for the set of simulations with a fixed number of physical modes. We show the $L_1$ error calculated using \eqref{eq:l1Herror_raw} for both constraints. The top two panels show the $L_1$ error for the initial conditions, at $z=1100$, and the bottom two panels show the $L_1$ error for $z=0$.}
    \label{fig:10dx_HMconverge_z0_z1100}
\end{figure}

\end{centering}

\section{Constraint violation} \label{appx:constraints}

In numerical relativity, the error can be quantified by analysing the violation in the Hamiltonian and momentum constraint equations, defined by
\begin{subequations} \label{eqs:HamMom}
	\begin{align}
		H &\equiv R - K_{ij}K^{ij} + K^{2} - 16\pi\rho = 0, \label{eq:ham} \\
		M_{i} &\equiv \tilde{\nabla}_{j} K^{j}_{\phantom{j}i} - \tilde{\nabla}_{i}K - 8\pi S_{i} = 0, \label{eq:mom}
	\end{align}
\end{subequations}
where $S_{i}=h_{i\alpha}n_{\beta}T^{\alpha\beta}$ and $\tilde{\nabla}_{i}$ represents the covariant derivative associated with the 3-metric $\gamma_{ij}$, and we define the magnitude of the momentum constraint to be $M=\sqrt{M_i M^i}$. An exact solution to Einstein's equations will identically satisfy \eqref{eqs:HamMom}. Since we are solving Einstein's equations numerically, we expect some non-zero violation in the constraints. We use the \texttt{mescaline} code, described in Section~\ref{subsec:mescaline}, to calculate the constraint violation as a function of time.

For the simulations we present in this work, we do not expect (in general) to see convergence of the constraint violation. At each different resolution we are sampling a different section of the power spectrum, and hence a different physical problem. In order to see convergence of the constraints at the correct order, we must analyse a controlled case in which the gradients are kept constant between resolutions. We perform three simulations at resolutions $32^3,64^3$, and $128^3$ inside an $L=1$ Gpc domain. We generate the initial conditions for the $32^3$ simulation using CAMB; restricting the minimum sampling wavelength to be $\lambda_{\rm min}=10\Delta x_{32} = 312.5$ Mpc. We use linear interpolation to generate the same initial conditions at $64^3$ and $128^3$. 

Figure~\ref{fig:constraints_control} shows the violation in the Hamiltonian (top panels) and momentum (bottom panels) constraints, for the set of simulations with a controlled number of physical modes, as a function of effective redshift. Left panels show the raw $L_1$ error for the violation, which for the Hamiltonian constraint we define as
\begin{equation}\label{eq:l1Herror_raw}
	L_1(H) = \frac{1}{n}\sum_{a=1}^{n} |H_a|,
\end{equation}
where $n$ is the total number of grid cells, and $H_a$ is the Hamiltonian constraint violation at grid cell $a$, and similarly for the momentum constraint. To quantify the "smallness" of this violation, we normalise the constraint violations to their relative "energy scales". Similar to \citep{mertens2016,giblin2017a}, we define
\begin{subequations} \label{eqs:energyscale}
	\begin{align}
		[H] &\equiv \sqrt{R^{2} + (K_{ij}K^{ij})^{2} + (K^{2})^{2} + (16\pi\rho)^{2}}, \label{eq:enham} \\
		[M] &\equiv \sqrt{(\tilde{\nabla}_{j} K^{j}_{\phantom{j}i})(\tilde{\nabla}_{k} K^{ki}) + (\tilde{\nabla}_{i}K)(\tilde{\nabla}^{i}K) + (8\pi)^2 S_i S^i}.\label{eq:enmom}
	\end{align}
\end{subequations}
Right panels in Figure~\ref{fig:constraints_control} show the relative $L_1$ error for each constraint violation, which we define as
\begin{equation}\label{eq:l1Herror_rel}
	L_1(H/[H]) = \frac{ \frac{1}{n}\sum_{a=1}^{n} |H_a| }{ \frac{1}{n} \sum_{a=1}^{n} [H]_a},
\end{equation}
where $[H]_a$ is the energy scale calculated at grid cell $a$, and similarly for the momentum constraint. We take the positive root of both $[H]_a$ and $[M]_a$.

Figure~\ref{fig:10dx_HMconverge_z0_z1100} shows the raw $L_1$ error for the same set of simulations, for the initial data ($z=1100$; top two panels) and for the data at $z=0$ (bottom two panels). The left panels show the $L_1$ error for the Hamiltonian constraint, and the right panels show the $L_1$ error for the momentum constraint. We see the expected second order convergence for the Einstein Toolkit, with the exception of the $N=128$ simulation's violation in the momentum constraint. Our initial speculation was that this was roundoff error, given the smallness of the quantities involved. The top panels of Figure~\ref{fig:10dx_HMconverge_z0_z1100} show that this issue is not due to the non-convergence of our initial data. Whether or not this is roundoff error remains unclear, and cannot be clarified without re-performing our simulations in quad precision.


For the simulations with a controlled number of modes, at $z=0$ for resolution $N=128$ the relative Hamiltonian constraint violation is $L_1(H/[H]) = 1.3\times10^{-3}$, the momentum constraint is $L_1(M/[M]) = 2.3\times10^{-2}$. For the simulations with full power spectrum sampling, at $z=0$ and resolution $N=256$ we find $L_1(H/[H]) = 4.4\times10^{-1}$, and $L_1(M/[M]) = 5.4\times10^{-2}$.

\begin{figure}
     \includegraphics[width=0.5\textwidth]{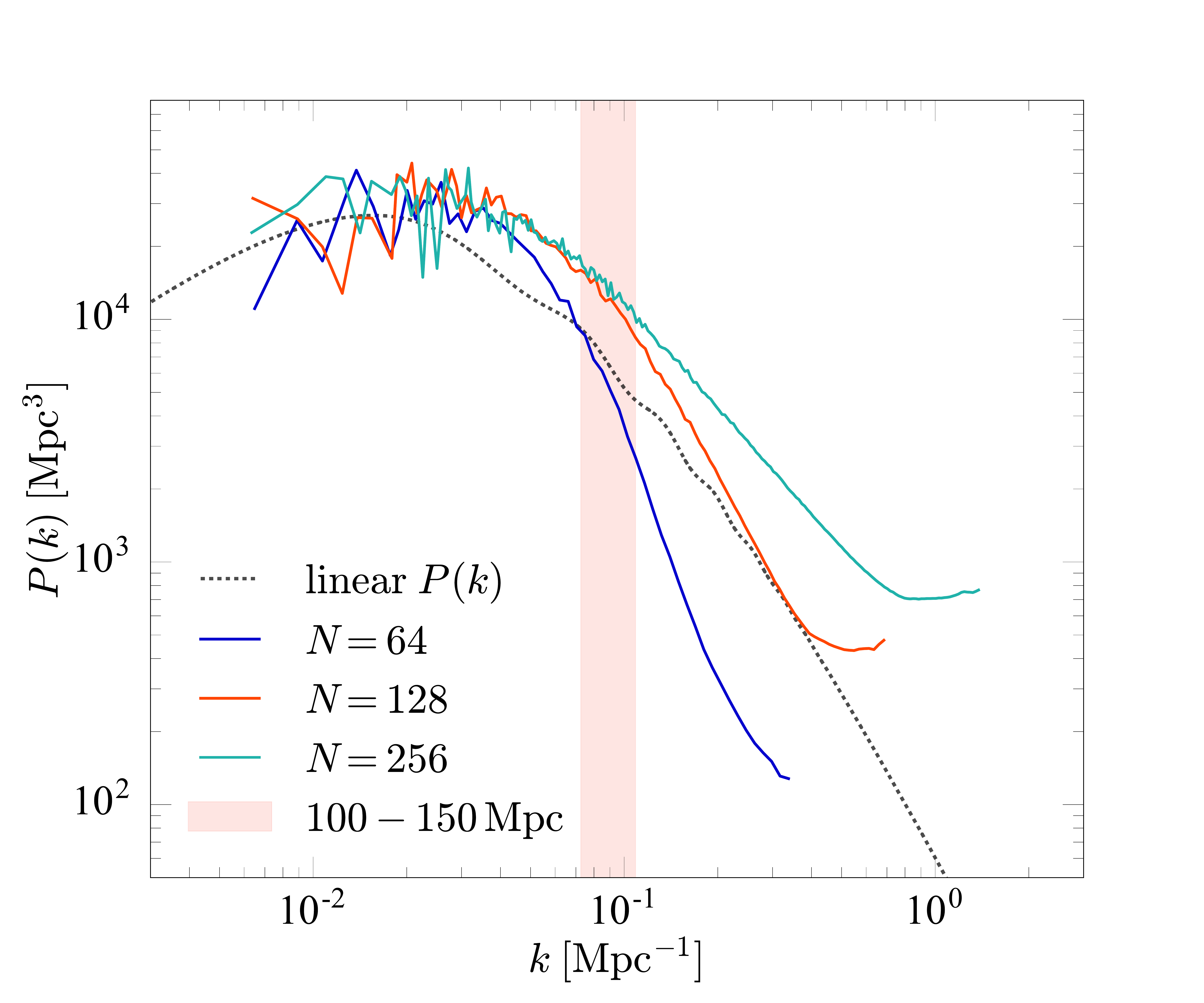}
    \caption{Power spectrum of fractional density fluctuations $\delta$ at $z=0$. Solid coloured curves show $P(k)$ for three simulations with $64^3,128^3,256^3$, each for an $L=1$ Gpc domain, and the dashed curve shows the linear power spectrum at $z=0$. The pink shaded region represents one-dimensional scales of $80-100$ Mpc. }
    \label{fig:pkz0}
\end{figure}


\section{Convergence and errors} \label{appx:convergence}
\begin{figure*}
\centering
\begin{subfigure}{0.5\textwidth}
  \centering
  \includegraphics[width=0.9\textwidth]{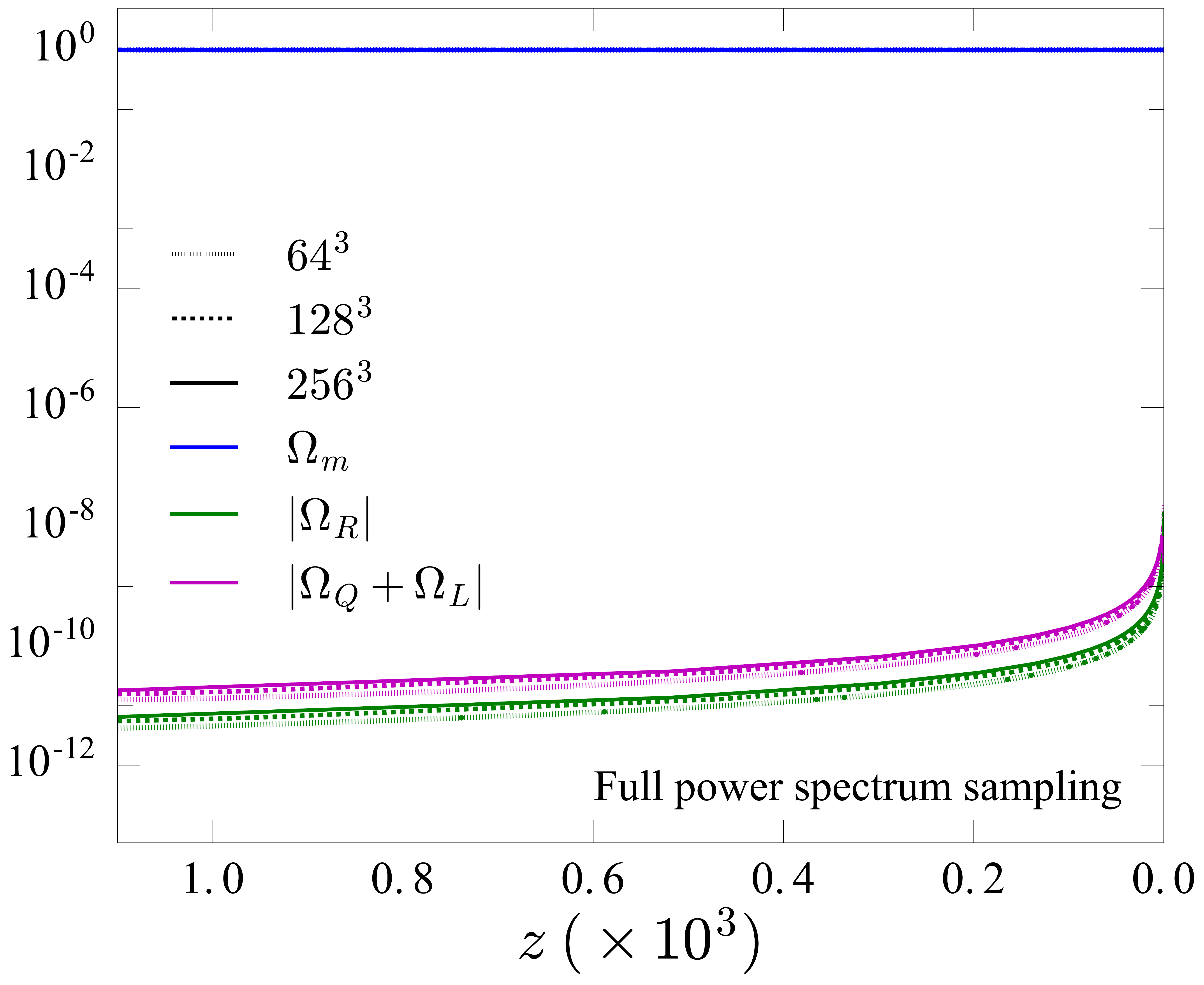}
  \label{fig:omegas_z_noncontrol}
\end{subfigure}%
\begin{subfigure}{0.5\textwidth}
  \centering
  \includegraphics[width=0.9\textwidth]{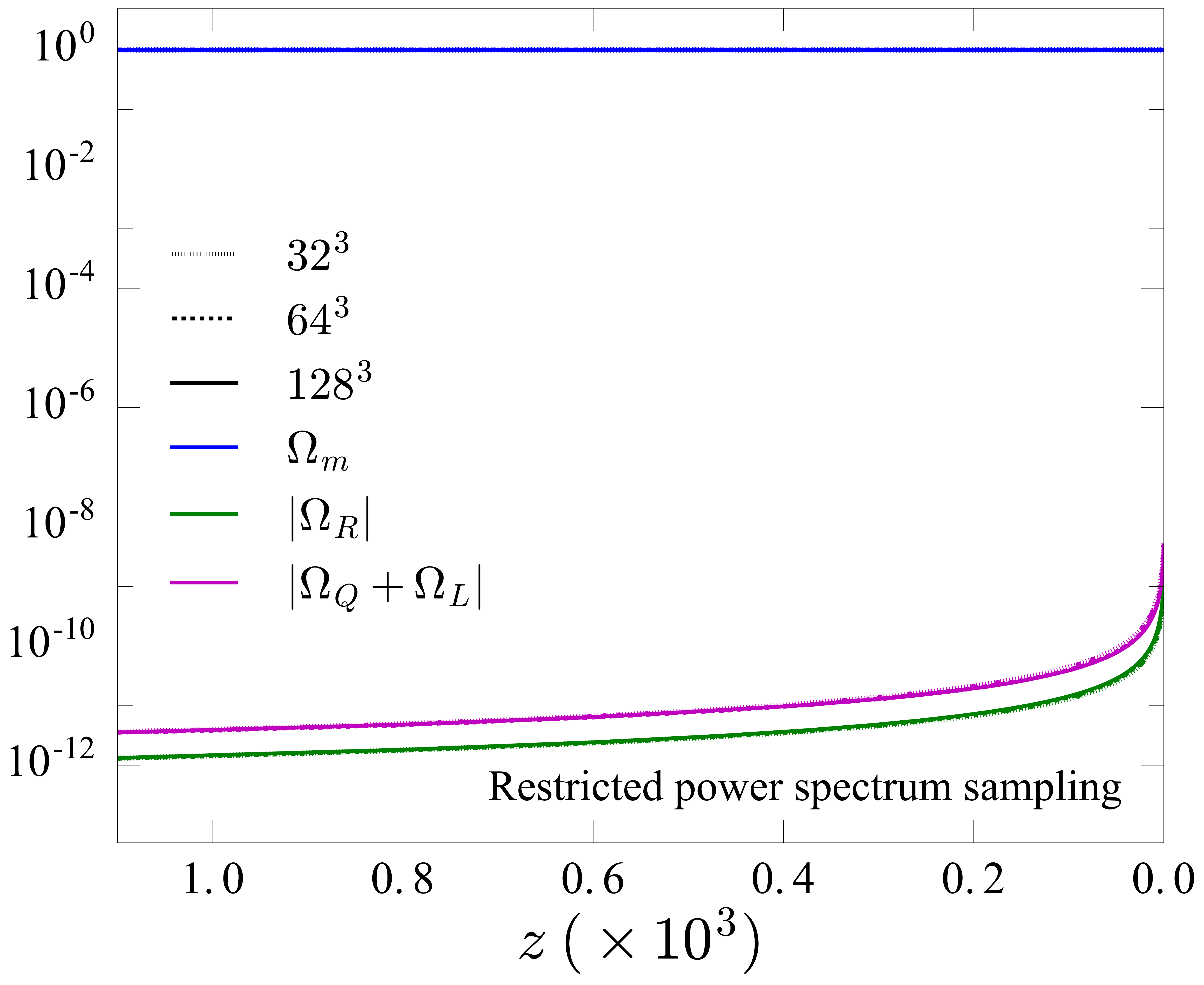}
   \label{fig:omegas_z_control}
\end{subfigure}
\caption{Global cosmological parameters as a function of effective redshift, for simulations with full power spectrum sampling (left panel) and a controlled number of physical modes (right panel). Dotted, dashed, and solid curves show resolutions as indicated in each seperate legend. Blue curves show $\Omega_m$, green curves show $|\Omega_R|$, and purple curves show $|\Omega_Q + \Omega_L|$.}
 \label{fig:omegas_z}
\end{figure*}

\begin{figure*}
     \includegraphics[width=0.9\textwidth]{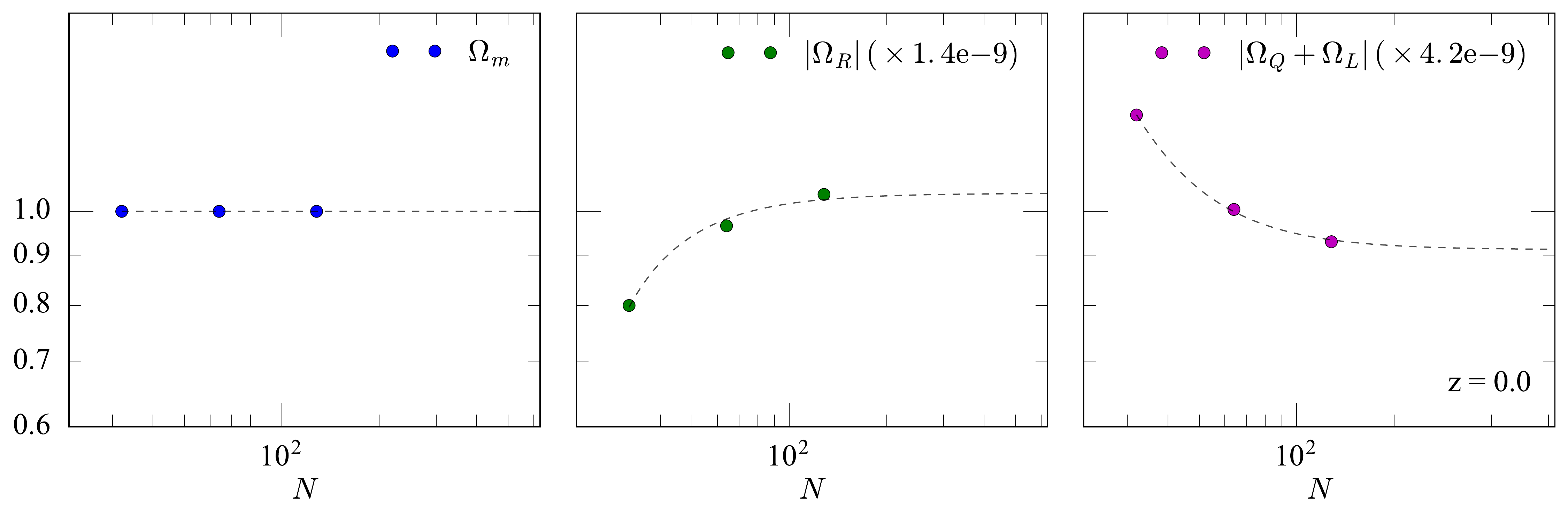}
    \caption{Cosmological parameters measured at $z=0$ for simulations with a controlled number of physical modes. Coloured points show $\Omega_m$ (left), $|\Omega_R|$ (middle), and $|\Omega_Q+\Omega_L|$ (right) for resolutions $N=32, 64,$ and $128$. Dashed curves are the convergence fit for each parameter, detailed in the text. We use these curves for a Richardson extrapolation to calculate the true value of the parameters and hence the errors on our measurements. }
    \label{fig:omegas_curvefit}
\end{figure*}

\begin{figure}[h]
     \includegraphics[width=0.45\textwidth]{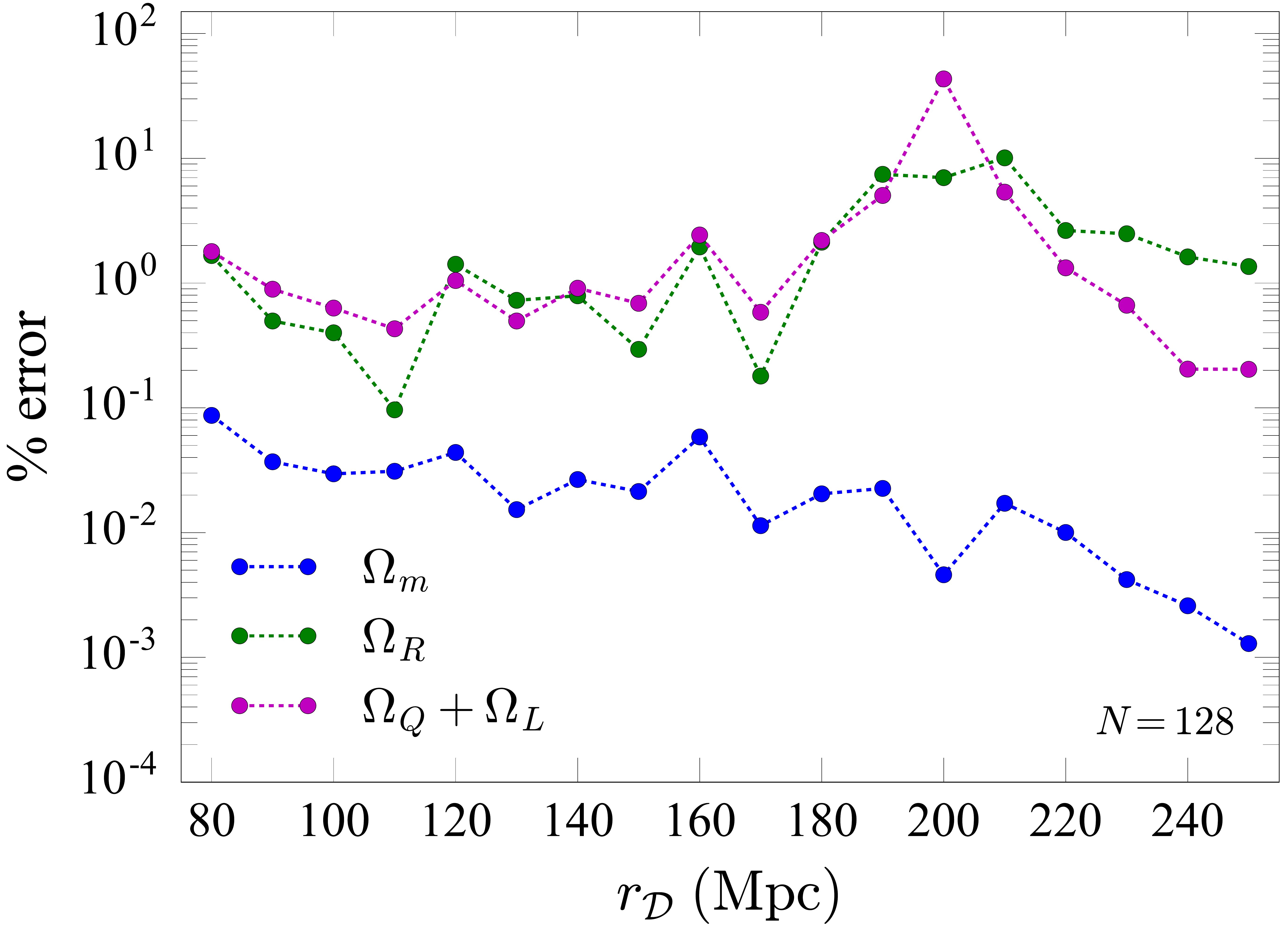}
    \caption{Richardson extrapolated errors for cosmological parameters calculated within subdomains, for the $128^3$ resolution controlled simulation, as a function of averaging radius $r_\mathcal{D}$. Blue points show the percentage error for $\Omega_m$, green points for $\Omega_R$, and purple points for $\Omega_Q+\Omega_L$.}
    \label{fig:omegas_rD_errors}
\end{figure}

In the previous section we discussed the convergence of the constraint violation, which for the main simulations presented in this paper we do not expect to reduce with resolution, since the physical problem is changing. Regardless of this, we expect the power spectrum of fractional density fluctuations, $\delta$, to be converged in these simulations. Coloured curves in Figure~\ref{fig:pkz0} show the $z=0$ power spectrum of the fractional density fluctuations for resolutions $64^3,128^3$, and $256^3$, each within an $L=1$ Gpc domain. The dashed curve shows the linear power spectrum at $z=0$ calculated with CAMB. The pink shaded region shows one-dimensional scales of $100-150$ Mpc, roughly corresponding to scales which are converged. Below these scales we therefore underestimate the growth of structures, and hence underestimate the contribution from curvature and backreaction in our calculations. 

Figure~\ref{fig:omegas_z} shows the global cosmological parameters as a function of redshift for simulations sampling the full power spectrum (left panel), and for those with a restricted sampling of the power spectrum; our controlled case discussed in the previous section (right panel). Dotted, dashed, and solid curves show different resolutions as indicated in each seperate legend. Blue curves show the density parameter $\Omega_m$, green curves show the curvature parameter $|\Omega_R|$, and purple curves show the backreaction parameters $|\Omega_Q+\Omega_L|$. As resolution increases in the left panel --- as we add more small-scale structure --- the contributions from curvature and backreaction increase, however still remain negligible relative to the matter content, $\Omega_m$, and are unlikely to grow large enough to be significant when reaching a realistic resolution. In the right panel, we have kept the physical problem constant and varied only the computational resolution, and so we see all global parameters converged towards a single value. Comparing the left panel to the right panel, the value of the curvature and backreaction parameters differ by almost an order of magnitude. This is due to the restricted power spectrum sampling for the simulations in the right panel, in which we only sample structures down to $\lambda_{\rm min}=10\Delta x_{32} = 312.5$ Mpc. These simulations should therefore not be considered the most realistic representation of our Universe. From this comparison we see that adding more structure results (in general) in a larger contribution from curvature and backreaction. 

We calculate the errors in the cosmological parameters using a Richardson extrapolation, which requires the gradients between resolutions to remain the same. This is not the case for the simulations with full power spectrum sampling, however it is the case for the controlled simulations with a restricted mode sampling. We therefore use the controlled simulations to approximate the errors for our main calculations. Figure~\ref{fig:omegas_curvefit} shows the values of the globally averaged cosmological parameters at $z=0$ for the controlled simulations. Coloured points show $\Omega_m$ (left panel), $|\Omega_R|$ (middle panel), and $|\Omega_Q+\Omega_L|$ (right panel) at resolutions $N=32,64,$ and $128$. We use the function \texttt{curve\_fit} as a part of the SciPy\footnote{https://scipy.org} Python package to fit each set of points with a curve of the form $\Omega_i(N) = \Omega_{\rm inf} + E\times N^{-2}$, where $\Omega_{\rm inf}$ is the value of the relevant cosmological parameter at $N\rightarrow\infty$, and $E$ is a constant. Black dashed curves in each panel of Figure~\ref{fig:omegas_curvefit} show the best-fit curves. 


The best-fit value for $\Omega_{\rm inf}$ provides an approximation of the correct value of each cosmological parameter for this set of test simulations. The residual between our calculations and $\Omega_{\rm inf}$ gives the error in our calculations. For the controlled simulation with $128^3$ resolution, the errors in the global cosmological parameters are $10^{-8}$, $4\times10^{-12}$, and $7\times10^{-11}$ for $\Omega_m$, $\Omega_R$, and $\Omega_Q+\Omega_L$, respectively. Expressed as a percentage error, these are  $10^{-6}\%$, 0.27\%, and 1.9\%.

We follow the same procedure to estimate the errors on the cosmological parameters calculated within subdomains. Figure~\ref{fig:omegas_rD_errors} shows the percentage error in each parameter as a function of averaging radius of the subdomain, $r_\mathcal{D}$, for the controlled simulation with $128^3$ resolution. Blue points show the error for $\Omega_m$, green points show $\Omega_R$, and purple points show $\Omega_Q+\Omega_L$. The jump in errors evident at $\sim 200$ Mpc in $\Omega_R$ and $\Omega_Q+\Omega_L$ is due to a change in sign of the curvature and backreaction parameters.

\bibliography{litreview}

\end{document}